\RequirePackage{graphicx}
\documentclass[aps,pra,floatfix,preprint, longbibliography]{revtex4-2}
\usepackage{amssymb}
\usepackage{enumitem}
\usepackage{mathtools}
\usepackage{xcolor}
\usepackage{float}
\usepackage{graphicx}
\usepackage{natbib}
\usepackage{dcolumn}
\usepackage{bm}
\usepackage{mathtools}
\usepackage{braket}
\usepackage[utf8x]{inputenc}
\usepackage{amsmath}
\usepackage{multirow}
\usepackage{amsfonts}
\usepackage{changepage}
\usepackage{amssymb}
\usepackage{rotating} 

\usepackage{amsthm}

\theoremstyle{plain}
\theoremstyle{definition}

\newcommand{\qGam}[4]{\hat{\sigma}^\dagger_{#1} \hat{\sigma}^\dagger_{#2} \hat{\sigma}^{}_{#3} \hat{\sigma}^{}_{#4}}
\newcommand{\fGam}[4]{\hat{a}^\dagger_{#1} \hat{a}^\dagger_{#2} \hat{a}^{}_{#3} \hat{a}^{}_{#4}}

\begin{document}

\title{Many-Fermion Simulation from the Contracted Quantum Eigensolver without Fermionic Encoding of the Wave Function}



\author{Scott E. Smart and David A. Mazziotti}
\email[]{damazz@uchicago.edu}
\affiliation{Department of Chemistry and The James Franck Institute, The University of Chicago, Chicago, IL 60637}%
\date{Submitted March 4, 2022}
%


\begin{abstract}
Quantum computers potentially have an exponential advantage over classical computers for the quantum simulation of many-fermion quantum systems.  Nonetheless, fermions are more expensive to simulate than bosons due to the fermionic encoding---a mapping by which the qubits are encoded with fermion statistics.  Here we generalize the contracted quantum eigensolver (CQE) to avoid fermionic encoding of the wave function.  In contrast to the variational quantum eigensolver, the CQE solves for a many-fermion stationary state by minimizing the contraction (projection) of the Schr{\"o}dinger equation onto two fermions.  We avoid fermionic encoding of the wave function by contracting the Schr{\"o}dinger equation onto an unencoded pair of particles.  Solution of the resulting contracted equation by a series of unencoded two-body exponential transformations generates an unencoded wave function from which the energy and two-fermion reduced density matrix (2-RDM) can be computed.  We apply the unencoded and the encoded CQE algorithms to the hydrogen fluoride molecule, the dissociation of oxygen O$_{2}$, and a series of hydrogen chains.   Both algorithms show comparable convergence towards the exact ground-state energies and 2-RDMs, but the unencoded algorithm has computational advantages in terms of state preparation and tomography.
\end{abstract}

\baselineskip24pt

\maketitle

\section{Introduction}

Simulations on quantum computers have a potentially exponential advantage for the computation of many-fermion quantum systems such as molecules and materials \cite{Abrams1997, Whitfield2011}. However, if each qubit represents the particle filling of an orbital, as in the formalism of second quantization,  the natural particles for simulation on a quantum computer---qubit particles---are hard-core bosons rather than fermions~\cite{Mazziotti2021,Wu2002}. Consequently, as originally recognized by Feynman \cite{Feynman1982}, the simulation of a many-fermion quantum system is potentially more complicated than the simulation of an equivalent many-boson quantum system.  The particle statistics of fermions are typically encoded in the qubit wave function in a process known as fermionic encoding, which increases computational complexity in terms of the quantum state preparation and tomography \cite{Jordan1928,Abrams1997,Bravyi2000,Seeley2012,Hastings2014, Wecker2014}.

To avoid this additional complexity, hardware-efficient wave functions have been developed for fermionic systems in which a wave function is prepared on the quantum computer such that its simulated particles are neither bosons nor fermions \cite{Kandala2017, Choquette2020, Barron2020}.  Encoding the Hamiltonian with fermion statistics in these instances still recovers the many-fermion energy from the arbitrary statistics of the prepared wave function.  While the absence of particle statistics can produce optimization difficulties such as barren plateaus, more accurate results have recently been obtained by using qubit-particle wave functions whose particles have the statistics of hard-core bosons \cite{Xia2020, Izmaylov2019, Ryabinkin2018, Ryabinkin2020, Ryabinkin2021, Tang2019}.  We have shown that, in contrast to the hardware-efficient wave functions, the qubit-particle wave functions are isomorphic to the fermion wave functions, and hence, they uniquely parameterize the set of ground-state two-fermion reduced density matrices (2-RDMs)~\cite{Mazziotti2021}. Consequently, the energy and 2-RDM of a many-fermion quantum system can be computed from a qubit-particle wave function with the accuracy associated with a fermion wave function but at a potentially reduced computational cost.

Recently, we presented a hybrid quantum-classical algorithm for the many-fermion problem known as the contracted quantum eigensolver (CQE)~\cite{Smart2021_prl,Boyn2021,Smart2022_benzyne}.  The CQE minimizes the residual of a contraction (projection) of the Schr{\"o}dinger equation onto the space of two particles.   The algorithm updates the fermion wave function iteratively with two-body exponential transformations to minimize the residual.   To keep the transformations unitary, we use the anti-Hermitian part of the contraction of the Schr{\"o}dinger equation, known as the anti-Hermitian contracted Schr{\"o}dinger equation (ACSE)~\cite{Mazziotti2006,Mazziotti2007, Mazziotti2007_mr,Gidofalvi2007,Rothman2009,Snyder2012,Sand2015,Boyn2021_spin}.   In contrast to the variational quantum eigensolver (VQE)~\cite{Peruzzo2014, McClean2016,Romero2017} in which the variational form of the wave function is not specified, the CQE produces a compact wave function Ansatz consisting of a series of two-body exponential transformations applied to the reference wave function. This CQE Ansatz~\cite{Mazziotti2007}, which is significantly more flexible than a truncated coupled cluster Ansatz, can be converged to the exact solution of the Schr{\"o}dinger equation.
The CQE algorithm, which stores just the 2-RDM on the classical computer, has a potentially exponential advantage over classical methods for solving the $N$-fermion problem like full configuration interaction.

Here we develop a generalization of the CQE algorithm for the many-fermion problem that solves an unencoded ACSE in which the anti-Hermitian part of the Schr{\"o}dinger equation is contracted onto two qubit particles rather than two fermions.  The generalized algorithm solves for the fermionic ground-state energy and 2-RDM by updating a qubit-particle wave function at each iteration with a two-qubit-particle unitary transformation that minimizes the residual of the unencoded ACSE.   We explore the accuracy and efficiency of the proposed algorithm through quantum simulations of the hydrogen fluoride molecule, the dissociation of diatomic oxygen O$_{2}$, and a series of hydrogen chains.  Both the encoded (fermion) and the unencoded (qubit-particle) CQE algorithms show similar convergence to the exact ground-state energies and 2-RDMs, but the unencoded CQE has potentially important computational savings in terms of the number of two-qubit gates required in the state preparation and the locality of the 2-RDM tomography.

\section{Theory}

We review the ACSE and its CQE algorithm for quantum simulation in section~\ref{sec:eACSE}, present the unencoded ACSE and its CQE algorithm that avoid fermionic encoding of the wave function in section~\ref{sec:uACSE}, explore the connection between the encoded and unencoded ACSEs in section~\ref{sec:euACSE}, and discuss practical considerations for both encoded and unencoded CQE algorithm in section~\ref{sec:practical}.

\subsection{Encoded ACSE and its CQE algorithm} \label{sec:eACSE}
Consider a fermionic quantum system of $N$ fermions in $r$ orbitals described by the Schr{\"o}dinger equation
\begin{equation} \label{eq:Ham}
{\hat H} | \Psi \rangle = E | \Psi \rangle.
\end{equation}
Here $E$ and $| \Psi_{n} \rangle$ are the many-fermion ground-state energy and wave function, and ${\hat H}$ is the Hamiltonian operator
\begin{equation}
{\hat H} = \sum_{pqst}{ ^{2} K^{pq}_{st} \fGam{p}{q}{t}{s} }
\end{equation}
in which $^{2} K$ is the reduced Hamiltonian matrix, the indices ranging from one to $r$ denote the orbitals, and ${\hat a}^{\dagger}_{i}$ and ${\hat a}_{i}$ are the creation and annihilation operators of the fermion in the $i^{\rm th}$ orbital. The ACSE is the anti-Hermitian contraction of the many-fermion Schr{\"o}dinger equation onto two fermions~\cite{Mazziotti2006,Mazziotti2007, Mazziotti2007_mr,Gidofalvi2007,Rothman2009,Snyder2012,Sand2015,Boyn2021_spin}:
\begin{equation}\label{eq:ACSE}
\langle \Psi | [ \fGam{i}{k}{l}{j},\hat{H} ] | \Psi \rangle = 0.
\end{equation}
As shown in previous work, iterative solution of the ACSE generates a unitary two-body exponential Ansatz for the wave function~\cite{Mazziotti2007, Smart2021_prl}
\begin{equation}
{\rm e}^{{\hat A}_{m}^{\rm F}} \cdots {\rm e}^{{\hat A}_{2}^{\rm F}} {\rm e}^{{\hat A}_{1}^{\rm F}}| \Psi_{0} \rangle
\end{equation}
in which $| \Psi_{n}^{0} \rangle$ is the reference wave function and the unitary transformation at the $m^{\rm th}$ iteration is determined by a two-body anti-Hermitian operator
\begin{equation} \label{eq:AFop}
{\hat A}_{m}^{\rm F} = \epsilon_{m} \sum_{ijkl}{ ^{2}_{\rm F} A^{ij;kl}_{m} \fGam{i}{j}{l}{k} }
\end{equation}
that corresponds to the residual of the ACSE
\begin{equation}\label{eq:AF}
{}^2_{\rm F} A^{ij;kl}_{m} = \langle \Psi_{m-1} | [\hat{H},\fGam{i}{j}{l}{k}]| \Psi_{m-1} \rangle .
\end{equation}
The residual the ACSE at the $m^{\rm th}$  iteration equals the gradient of the energy with respect to the two-body anti-Hermitian operator $ {\hat A}_{m}^{\rm F}$.  Hence, the residual the ACSE is zero not only when the ACSE is satisfied but also when the gradient of the energy vanishes.  The $\epsilon_{m} $ is a step-like parameter that can be optimized at the $m^{\rm th}$ iteration to minimize the energy.  From the wave function the elements of the 2-RDM at the $m^{\rm th}$ iteration can be computed
\begin{equation} \label{eq:D2}
^{2} D^{pq;st}_{m} = \langle \Psi | \fGam{p}{q}{t}{s} | \Psi \rangle  .
\end{equation}
While calculation of the residual the ACSE and the 2-RDM on the classical computer typically requires a cumulant approximation for the three-particle reduced density matrix (3-RDM)~\cite{Mazziotti1998, Mazziotti1998_schwinger} to avoid storage of the wave function, both the ACSE residual and the 2-RDM in the CQE algorithm can be directly calculated by quantum tomography. Implementation of the state preparation and tomography in the CQE requires fermionic encoding in which the fermionic creation and annihilation operators are expressed in terms of qubit operators through a transformation such as the Jordan-Wigner mapping.

\subsection{Unencoded ACSE and its CQE algorithm} \label{sec:uACSE}

\begin{figure}
    \centering
    \includegraphics[scale=0.4]{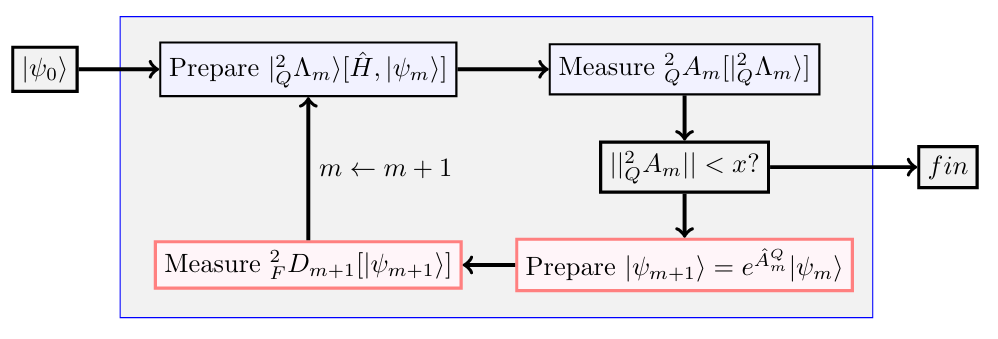}
    \caption{Unencoded CQE algorithm. Given some initial state that we can prepare on the quantum computer, we alternate between solving the unencoded ACSE on the quantum computer, and updating the new wave function given information from the unencoded ACSE.}
    \label{fig:ucqe}
\end{figure}

In this paper we generalize the CQE algorithm to solve the many-fermion problem using a qubit-particle wave function that does not require fermionic encoding.  Consider the anti-Hermitian contraction of the Schr{\"o}dinger equation onto two qubit particles to generate the unencoded ACSE
\begin{equation}\label{eq:qpACSE}
\langle \Psi | [ \qGam{i}{k}{l}{j},\hat{H} ] | \Psi \rangle = 0.
\end{equation}
where the Hamiltonian is defined with fermionic operators as in Eq.~(\ref{eq:Ham}) but the ${\hat \sigma}_{i}^{\dagger}$ and ${\hat \sigma}_{i}$ are the creation and annihilation operators of a qubit particle in the $i^{\rm th}$ orbital.  As in the previous case of the contraction onto two fermions to generate the ACSE (or encoded ACSE), iterative solution of the unencoded ACSE generates a unitary two-qubit-particle exponential Ansatz for the wave function
\begin{equation} \label{eq:expQ}
{\rm e}^{{\hat A}_{m}^{\rm Q}} \cdots {\rm e}^{{\hat A}_{2}^{\rm Q}} {\rm e}^{{\hat A}_{1}^{\rm Q}}| \Psi_{0} \rangle
\end{equation}
in which the unitary transformation at the $m^{\rm th}$ iteration is determined by a two-qubit-particle anti-Hermitian operator
\begin{equation}\label{eq:AQop}
{\hat A}_{m}^{\rm Q} = \epsilon_{m} \sum_{ijkl}{ ^{2}_{\rm Q} A^{ij;kl}_{m} \qGam{i}{j}{l}{k} }
\end{equation}
that corresponds to the residual of the unencoded ACSE
\begin{equation}\label{eq:AQ}
{}^2_{\rm Q} A^{ij;kl}_{m} = \langle \Psi_{m-1} | [\hat{H},\qGam{i}{j}{l}{k}]| \Psi_{m-1} \rangle .
\end{equation}
The residual of the unencoded ACSE at the $m^{\rm th}$ iteration equals the gradient of the energy with respect to the two-qubit-particle anti-Hermitian operator.  Computation of the 2-RDM uses the definition in Eq.~(\ref{eq:D2}).  Importantly, the CQE algorithm for solving the unencoded ACSE does not require fermionic encoding in the preparation of the wave function since the exponential Ansatz in Eq.~(\ref{eq:expQ}) is expressed entirely in terms of qubit-particle creation and annihilation operators.  Only the definitions of the 2-RDM and the Hamiltonian use fermionic second-quantized operators that require fermionic encoding into qubits for evaluation on quantum computers. A schematic of the CQE algorithm is shown in Fig.~\ref{fig:ucqe}.


Following previous work \cite{Smart2021_prl}, we can compute the residual of the ACSE through an auxiliary state
\begin{equation}
|{}^2_Q \Lambda_{m-1}\rangle = e^{-i\delta\hat{H}}|\Psi_{m-1}\rangle
\end{equation}
such that
\begin{equation}\label{eq:auxiliarysolution}
{}^2_Q A^{ij;kl}_m + \mathcal{O}(\delta^2) = \delta^{-1} \Im \left ( \langle \Lambda_{m-1}| \qGam{i}{j}{l}{k} |\Lambda_{m-1} \rangle \right ),
\end{equation}
where $\Im(x)$ is the imaginary component of $x$ and $\delta$ is a short time-like step, which for stochastic simulations should be larger than the sampling error.  If the RDM solutions are complex, we can evaluate the residual by a centered finite difference using two auxiliary states at $\pm \delta$~\cite{Smart2021_prl}. The approximation from a finite $\delta$ can potentially be minimized by using extrapolative techniques as in the unitary decomposition of operators~\cite{Schlimgen2021}. As discussed in the Appendix, the evaluation of the residual via the auxiliary state can be implemented efficiently in terms of the two-qubit gate count through first-order or Cholesky factorizations of the Hamiltonian.  For completeness we also note that it is possible to compute the residual without defining an auxiliary state but that the required tomography involves the measurement of a 4-particle RDM.

\subsection{Second-order Corrections to the Wave Functions and 2-RDMs} \label{sec:so}

Solution of either the encoded or unencoded ACSE can be accelerated through a second-order correction at each iteration. We have the following generalized exponential expansions
\begin{equation} \label{eq:expF2}
{\rm e}^{{\hat B}_{m}^{\rm F}} \cdots {\rm e}^{{\hat B}_{2}^{\rm F}} {\rm e}^{{\hat B}_{1}^{\rm F}}| \Psi_{0} \rangle
\end{equation}
and
\begin{equation} \label{eq:expQ2}
{\rm e}^{{\hat B}_{m}^{\rm Q}} \cdots {\rm e}^{{\hat B}_{2}^{\rm Q}} {\rm e}^{{\hat B}_{1}^{\rm Q}}| \Psi_{0} \rangle
\end{equation}
in which
\begin{equation}
{\hat B}_{m}^{\rm F} = {\hat A}_{m}^{\rm F} + {\hat C}_{m}^{\rm F}
\end{equation}
and
\begin{equation}
{\hat B}_{m}^{\rm Q} = {\hat A}_{m}^{\rm Q} + {\hat C}_{m}^{\rm Q}
\end{equation}
where either ${\hat C}_{m}^{\rm F}$ or ${\hat C}_{m}^{\rm Q}$ are modifications to the gradient direction and can be equal to (i) zero to recover the first-order Ansatz or (ii) a second-order correction such as that from a quasi-Newton method~\cite{Robinson2006}.  In general, a quasi-second-order correction accelerates convergence of the CQE, the implementation and further development of which is explored in concurrent work  \cite{Smart2022_opt}.

\subsection{Connection between the Encoded and Unencoded ACSEs} \label{sec:euACSE}

Before we consider the relationship of the solutions to the encoded ACSE, or just ACSE, and the unencoded ACSE, we examine the relationship between the solution of the ACSE and the solution of the many-fermion Schr{\"o}dinger equation. Consider the contracted Schr{\"o}dinger equation (CSE)~\cite{Mazziotti1998, Mazziotti2007_book, Nakatsuji1976}
\begin{equation}
\langle \Psi | \fGam{i}{j}{l}{k} \left ( {\hat H} – E \right ) | \Psi \rangle = 0 .
\end{equation}
If we expand the wave function in terms the eigenfunctions of the Hamiltonian
\begin{equation}
| \Psi \rangle = \sum_{n}{ c_{n} | \Psi_{n} \rangle },
\end{equation}
we can formally write the CSE as
\begin{equation}
\sum_{mn}{ c^{*}_{m} c_{n} E_{n} \langle \Psi_{m} | \fGam{i}{j}{l}{k} | \Psi_{n} \rangle } - E = 0.
\end{equation}
Because the CSE is equivalent to the energy variance which implies the Schr{\"o}dinger equation, the CSE is satisfied by an $N$-fermion wave function if and only if the Schr{\"o}dinger equation is satisfied~\cite{Mazziotti1998, Mazziotti2007_book, Nakatsuji1976}.  Hence, the CSE is zero for the ground state if and only if the expansion coefficients $c_n$ of excited states ($n>1$) vanish.   Substitution of the wave function expansion into the ACSE in Eq.~(\ref{eq:ACSE}) yields the expression
\begin{equation}
\sum_{mn}{ c^{*}_{m} c_{n} \left (E_{n}-E_{m} \right ) \langle \Psi_{m} | \fGam{i}{j}{l}{k} | \Psi_{n} \rangle }.
\end{equation}
As with the CSE, the ACSE is zero if the expansion coefficients $c_n$ of excited states ($n>1$) vanish.  This condition also implies the CSE as well as the Schr{\"o}dinger equation.  In contrast to the CSE, however, the ACSE does not strictly imply the Schr{\"o}dinger equation~\cite{Mazziotti2007}.  It is theoretically possible for the ACSE to vanish due to a cancellation of the anti-Hermitian terms, which in the expansion are represented by the energy differences $(E_{n}-E_{m})$.  Practical calculations, however, indicate that such cancellations do not occur easily and that the ACSE can in principle be converged to exact ground-state energies and 2-RDMs, especially with quasi-second-order corrections.

To understand the relationship between the solutions of the ACSE and unencoded ACSE, we express the residuals of both equations in terms of two parts
\begin{equation}\label{eq:AF2}
{}^2_{\rm F} A^{ij}_{kl} = \langle \Psi | [\hat{H},\fGam{i}{j}{l}{k}]| \Psi \rangle_{+} + \langle \Psi | [\hat{H},\fGam{i}{j}{l}{k}]| \Psi \rangle_{-}
\end{equation}
and
\begin{equation}\label{eq:AQ2}
{}^2_{\rm Q} A^{ij}_{kl} = \langle \Psi | [\hat{H},\qGam{i}{j}{l}{k}]| \Psi \rangle_{+} + \langle \Psi | [\hat{H},\qGam{i}{j}{l}{k}]| \Psi \rangle_{-}
\end{equation}
where the plus (minus) indicates the contributions from the wave function to the expectation value with net even (odd) permutations of particles.  Because the fermion and qubit-particle expectation values differ only from the sign of the odd permutations, we have the following two important relations
\begin{eqnarray}
\langle \Psi | [\hat{H},\qGam{i}{j}{l}{k}]| \Psi \rangle_{+} & = & + \langle \Psi | [\hat{H},\fGam{i}{j}{l}{k}]| \Psi \rangle_{+}  \\
\langle \Psi | [\hat{H},\qGam{i}{j}{l}{k}]| \Psi \rangle_{-} & = & - \langle \Psi | [\hat{H},\fGam{i}{j}{l}{k}]| \Psi \rangle_{-}  .
\end{eqnarray}
Substituting these relations into the residual of the unencoded ACSE yields the following equation
\begin{equation}
{}^2_{\rm Q} A^{ij}_{kl} = \langle \Psi | [\hat{H},\fGam{i}{j}{l}{k}]| \Psi \rangle_{+} - \langle \Psi | [\hat{H},\fGam{i}{j}{l}{k}]| \Psi \rangle_{-}
\end{equation}
Comparing this equation for the qubit-particle residual with the fermion residual in Eq.~(\ref{eq:AF2}), we observe that the only difference in the two residuals is the sign change of the second term.  If both the plus and minus terms converge to zero, then both the ACSE and unencoded ACSE produce identical solutions.  While it is in principle possible for the plus and minus terms to produce a spurious solution through an exact cancellation, because the Schr{\"o}dinger equation implies both the ACSE and the unencoded ACSE, both the positive and negative terms will tend to zero as the energy is minimized towards a stationary state by either encoded  or unencoded unitary transformations.  The relative magnitudes of the residuals $|| {}^2_{\rm F} A ||$ and $|| {}^2_{\rm Q} A ||$ are related to the relative rates of convergence of the ACSE and the unencoded ACSE.  A larger residual norm indicates less cancellation of the plus and minus terms which is likely to result in a faster rate of convergence towards the solution of the Schr{\"o}dinger equation.

\subsection{Practical Considerations of the CQE Algorithm} \label{sec:practical}

Finally, we introduce two practical aspects of the CQE , applicable to both encoded and unencoded variants, that are important for its implementation on a quantum computer.  We discuss: (i) a sparsification of the ${\hat A}_{m}$ operators in which at each iteration matrix elements below a given threshold are set to zero, and (ii) an approximate combination of ${\hat A}_{m-n}$ operators for $n \in [0,p]$ in which parts of the set of $p+1$ operators are combined to decrease the circuit length. Given the matrix elements of the residual of the encoded or unencoded ACSE in Eqs.~(\ref{eq:AF}) or~(\ref{eq:AQ}), respectively, we define a sparsification operation that zeros matrix elements below a given threshold
\begin{equation}
{}^{2} {\tilde A}^{ij;kl}_{m} = \textsc{sparse}[c]  ({}^{2} { A}^{ij;kl}_{m}) =
\begin{cases} 0  ~&\text{if} ~ | ^{2} A^{ij;kl}_{m}| < c|| ^{2} A_{m} ||_\infty  \\
^{2} A^{ij;kl}_{m} &\text{if} ~ |^{2} A^{ij;kl}_{m}| \geq c|| ^{2} A_{m} ||_\infty
\end{cases},
\end{equation}
where the scalar factor $c \in [0,1]$ and  the infinity norm $||A||_\infty $ of $A$ is the element with the largest absolute value.  When $c=0$, $\textsc{sparse}[0]$ is equivalent to the identity mapping, and when $c=1$, $\textsc{sparse}[1]$ is equivalent to selecting only the highest amplitude element.   Given a choice of the parameter $c$, we use this mapping at each iteration of the CQE to prune the residual matrix $^{2} A_{m}$.   Formally, at each iteration $^{2} A_{m}$ is replaced by $^{2} {\tilde A}_{m}$.

After defining a sparser matrix $^{2} {\tilde A}_{m}$ in $\hat{A}_{m}$, we still need to express the operator ${\hat A}_{m}$ as a product of unitary transformations, which is traditionally performed by trotterization.  We choose a first-order trotterization which is valid in the case that $||{}^2 {\tilde A}_{m}||$ is not too large.   Moreover, because the algorithm is greedy by design with the gradient being used at each iteration, the algorithm has the ability to adjust itself in part to errors in previous iterations including those from trotterization.

To address the growth of the circuit length with iterations, we define an approximate combination of ${\hat A}_{m-n}$ operators for $n \in [0,p]$, which we call the $p$-depth.  When updating the wave function at the $m^{\rm th}$ iteration, we examine the elements of  $^{2} A^{ij;kl}_{m}$ that were not pruned in one of the $p$ previous steps.  If $^{2} {\tilde A}^{ij;kl}_{m-n}$ for an $n$ in $[0,p]$ is non-zero, we update that term as follows:
\begin{equation}
^{2} {\tilde A}_{m-n}^{ij;kl} \leftarrow {}^{2} {\tilde A}_{m-n}^{ij;kl} + ^{2} A_m^{ijkl}.
\end{equation}
In this manner we collect terms to decrease circuit length even if the collection is approximate.  Importantly, the computed residual of the ACSE at the next iteration adjusts for errors introduced in previous terms in the product expansion of the wave function.  Other works exploit a classification of the commutating operators in the Hamiltonian or the unitary transformation of the wave function to minimize the circuit length, and these schemes can also be applied to the CQE algorithm.

The $p$-depth and $\textsc{sparse}[c]$ techniques are closely related, and in their extreme limits they produce specific Ans\"atze.  For $c=1$, a large $p$ is reasonable, because each iteration will contain only one or a few terms, and hence, setting $p=1$ or $p=2$ will not simplify the product of terms. On the other hand, when $c=0$, setting $p=1$ or $p=2$ may be necessary because a larger $p$ may generate a single exponential Ansatz, which negates a critical benefit of the CQE---the ACSE Ansatz for the wave function.

\section{Applications}

\subsection{Molecular Simulations}

\begin{figure}[h]
\includegraphics[scale=0.9]{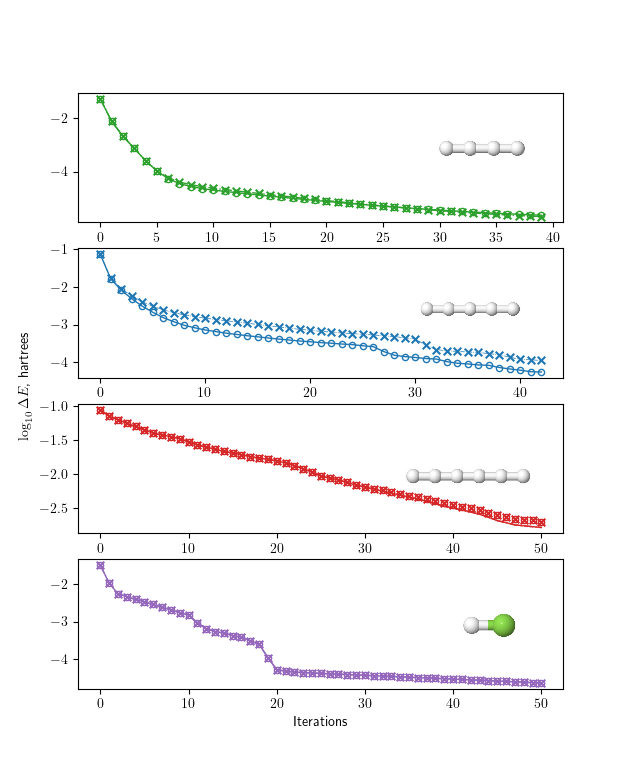}

\caption{Depictions of the quantum ACSE with an unencoded ${}^2_Q A_n $  ($\times$) and an encoded ${}^2 A_n$ ($\circ$) for linear H$_4$, H$_5$, H$_6$ and hydrogen fluoride at equilibrium bond distance (in the STO-3G basis). For most systems, nearly identical rates of convergence are obtained with similar accuracy.} \label{fig:equilibrium}
\end{figure}

We compare the encoded and unencoded CQE for several molecular systems, H$_4$, H$_5$, H$_6$, as well as hydrogen fluoride, at equilibrium geometries in the minimal Slater-type orbital basis (STO-3G) set \cite{HehreW.J.;DitchfieldR.;Pople1972}.  The convergence of the two CQE algorithms is shown in Fig.~\ref{fig:equilibrium}. For H$_6$ we select $c=1$ in the truncation, whereas for the other cases $c=0.1$.  In all cases the encoded and unencoded algorithms show convergence towards the solution of the Schr{\"o}dinger equation in the given basis set even without any second-order acceleration of the transformations at each iteration.  In some instances, the two algorithms can exhibit nearly identical convergence, especially in systems with significant pairing of the orbitals, as in an antisymmetrized geminal power wave function~\cite{Coleman1997,Coleman2000,Johnson2013,Stein2014}, where the particle statistics become less important~\cite{Sager2022}.  One such example is hydrogen fluoride whose hole wave function in the minimal basis set is a single two-hole function or a geminal.  The molecule H$_{5}$ exhibits greater differences between the fermionic and qubit-particle Ans\"atze, which is expected for open-shell or strongly correlated molecules.

\begin{figure}[h]

\includegraphics[scale=0.8]{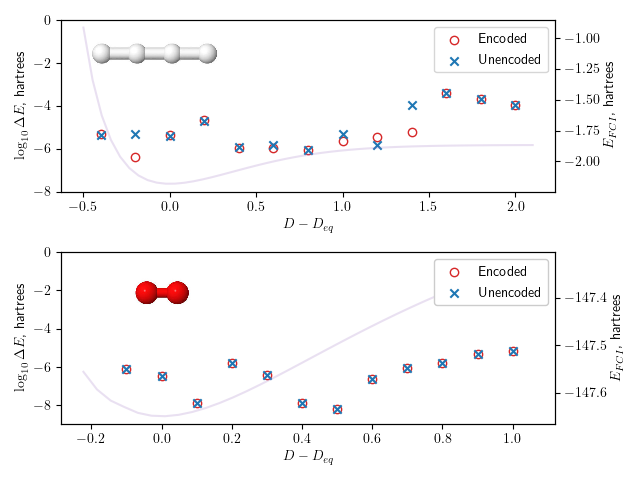}
\caption{Dissociation plots for 8 qubit simulations of H$_4$ and a 6-electron and 4-orbital active space of O$_2$ with the encoded and unencoded CQE. Convergence criteria is 0.001 for the norm of the ${}^2 B$ matrix which contains a correction to the ${}^{2} A$ matrix from a second-order approach\footnote{A quasi-Newton quantum CQE approach which adjusts the step direction based on a limited-memory Broyden-Fletcher-Goldarb-Shanno (BFGS) approach, where $2$ or $3$ vectors are stored. More details regarding this method will be presented in a subsequent work.}.  Both the encoded and unencoded CQE exhibit similar accuracy at both equilibrium and non-equilibrium geometries.} \label{fig:dissociaton}
\end{figure}

To examine performance for non-equilibrium geometries, we apply the encoded and unencoded CQE to computing the potential energy curves of H$_4$ and O$_2$.  We use approximate second-order transformations, based on a quasi-Newton method, as each iteration to accelerate convergence.  Figure~\ref{fig:dissociaton} shows the obtained energies across the dissociation curve. For a convergence criteria of 0.001 in the ACSE's residual norm, we consistently obtain high accuracy results at both equilibrium and non-equilibrium geometries regardless of the encoding.

\subsection{Investigation of Resource Requirements}

We examine the resource requirements for the encoded and unencoded methods.  Two cases, equilibrium and stretched geometries of linear H$_4$, are considered.  We examine the necessary resources to converge the residual of the ACSE within 0.01 as a function of the $p$-depth and the $\textsc{sparse}$ $c$ parameter with results in Table~\ref{tab:compare}.

First, in the equilibrium case (top half), for $c=1$, neither the fermionic nor the qubit-particle Ansatz changes with depth but each requires more iterations than when $c=0$ or $c=0.1$. In all of the instances, the qubit-particle wave function matches the fermionic wave function's trends but has 70$\pm$3\% of the fermionic CNOT cost. Additionally, the $p$-depth leads to a significant reduction in the number of terms. Including the previous iteration ($p=1$) results in 55\% of the $p=0$ count for $c=0$, and 66\% of the $p=0$ count for $c=0.1$ (equivalent circuits for $c=1$). The $p=3$ case shows even further reductions, requiring about 13\% and 34\% of the $p=0$ cost for the $c=0$ and $c=0.1$ cases, respectively. We can reasonably infer that the $p$-depth and  $\textsc{sparse}$ of the ${}^2 A$ operator can greatly impact the CQE's resource requirements as well as its rate of convergence.

For the stretched geometries the savings are more difficult to analyze.  For $c=1$ again we have a similar picture across $p$-depths, with notably more iterations required than for the equilibrium case. Again, the qubit case appears to have a similar advantage in the CNOT gate reduction versus the fermionic case. For other truncations, we lose the correspondence between the qubit-particle and fermionic wave functions. For multiple cases (i.e., when $(c,p)=(0,0),(0.1,3),(0.1,0)$), the qubit case requires either more iterations or yields a higher CNOT gate count. However, the shortest apparent Ansatz is still the qubit $c=0,~p=3$ case. The difference in the $p$-depth plays a more significant role, which leads to almost a 100-fold decrease in CNOT requirements between $p=3$ and $p=0$ for $c=0$, as well as a similar trend for $c=0.1$.

\begin{table}[h]
\label{tab:compare}
\caption{Comparing the $p$-depth and differing values of $\textsc{sparse}[c]$ (for ${}^2_F A $ and ${}^2_Q A$ in terms of their elements relative to the largest element) for two different lengths of H$_4$, in terms of the maximal CNOT gate count and number of iterations (in brackets, $[\cdot]$). A stopping criteria of 0.01 is selected for the gradient, and a quadratic trust-region step is used for choosing the step length at each step. For each instance, a similar accuracy is achieved in the fermionic and qubit cases, and for the most part, the qubit results displays a constant reduction in the number of necessary CNOT gates.}
\begin{tabular}{cc|ccc|c}
\multicolumn{6}{c}{$D=D_{eq}$} \\
\hline
\multirow{2}{*}{} & \multirow{1}{*}{$\textsc{sparse}[c]$} & \multicolumn{3}{c|}{$p-$depth} & \multirow{2}{*}{Average $\Delta E$} \\
&  & $p=3$ & $p=1$ & $p=0$ & \\
\hline
\multirow{3}{*}{${}^2_F \hat{A}_n $ } & $c=0 $ & 1568 [9] & 6782 [9]& 12494 [9] & $2 *10^{-5}$\\
& $c=\frac{1}{10}$ & 2020 [8] & 3858 [8]& 6068 [8] & $ 3 *10^{-5}$ \\
&  $c=1$  & 1263 [16]& 1408 [16] & 1408 [16] & $3 *10^{-5}$\\
\hline
\multirow{3}{*}{${}^2_Q \hat{A}_n $ }& $c=0  $  & 1162 [9] & 4974 [9]&  9014 [9] & $2 *10^{-5}$\\
& $c=\frac{1}{10}$  & 1414 [8] & 2756 [8]& 4098 [8] & $3 *10^{-5}$\\
& $c=1$ & 840 [16] & 928 [16] & 928 [16] & $4 *10^{-5}$\\
\hline
\multicolumn{6}{c}{$D=D_{eq}+1 $\AA } \\
\hline
\multirow{3}{*}{${}^2_F \hat{A}_n $ } &$c= 0 $ & 1558 [37] & 46146 [60]& 56058 [38] & $6 *10^{-4}$\\
& $c=\frac{1}{10}$ & 1574 [26] & 34190 [45]& 38706 [30] & $ 7 *10^{-4}$ \\
&  $c=1$& 10468 [146]& 12190 [161] & 12322 [161] & $2 *10^{-4}$\\
\hline
\multirow{3}{*}{${}^2_Q \hat{A}_n $ } & $c=0 $& 1162 [19] & 41992 [80]&  102540 [95] & $5 *10^{-4}$\\
& $c=\frac{1}{10}$ & 10938 [42] & 26840 [52]&  85324 [86] & $7 *10^{-4}$\\
&  $c=1$  & 6814 [155] & 7376 [141] & 7464 [141] & $3 *10^{-4}$\\
\hline

\end{tabular}
\end{table}

\begin{figure}[h]

\includegraphics[scale=0.95]{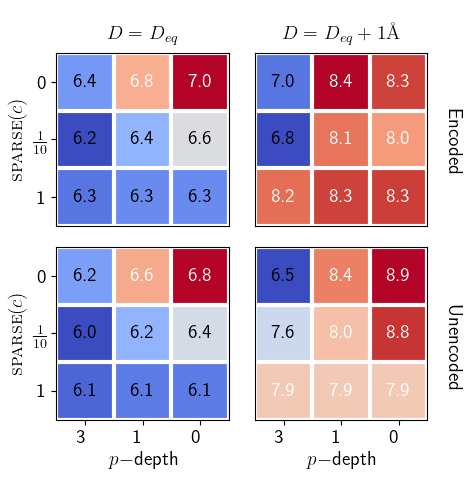}
\caption{Comparison of total resource count (given in $\log_{10}$ number of CNOT gates required) for equilibrium (left) and non-equilibrium (right) H$_4$ geometries, mirroring data seen in Table 1. The vertical axis contains information on whether encoded or unencoded operators are used, as well as the sparsification of ${}^2 A$. The horizontal axis indicates the $p$-depth, or the number of previous iterations in the Ansatz to which terms are added. The resource count also accounts for the number of fermionic energy and gradient evaluations.}
\label{fig:resource_comparison}
\end{figure}

While it is clear that low CNOT cases overall can be found with the $c=1$ instance, there is potentially a trade off with the number of iterations. Figure~\ref{fig:resource_comparison} explores the total resource count for the two H$_4$ geometries that accounts for the number of iterations, circuit measurements, and function and gradient evaluations. In the equilibrium case the lowest resource count for both the encoded and unencoded CQE is not $c=0$ but $c=0.1$ with $p=3$. The highest, on the other hand, is $c=0$ and $p=0$, which highlights the importance of simplifying the Ansatz.  Additionally, in all cases the unencoded Ansatz outperforms the encoded Ansatz.  The non-equilibrium geometry yields a similar picture to the equilibrium geometry, albeit at higher costs overall. The key difference arises for $c=0$ and $p=3$ where significant improvements visible. Results indicate that the qubit-particle Ansatz can require more resources than the fermionic one, and that care should be taken in choosing both $c$ and $p$.

\subsection{Generation of the ${}^2_Q A$ Matrix}

\begin{figure}[h]
\includegraphics[scale=0.75]{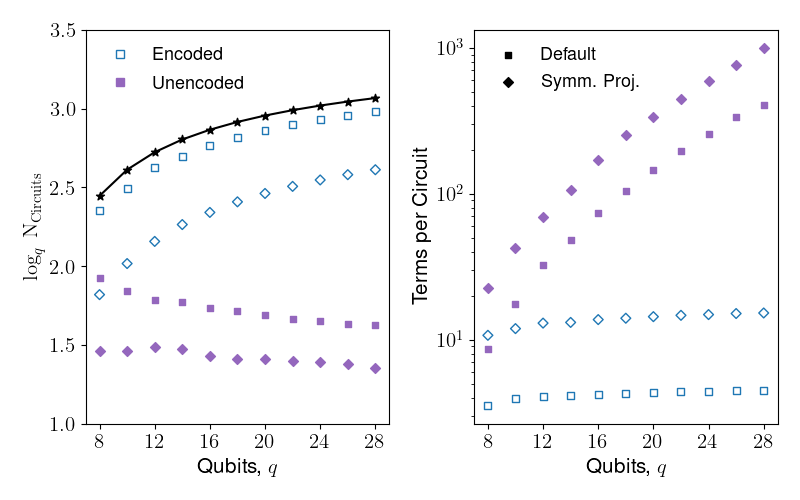}
\caption{(Left) Effective scaling of the tomographic complexity of the qubit-particle and fermionic ${}^2 A$ matrices (under the Jordan-Wigner transformation) with symmetry projection of $\hat{N}$ and $\hat{S}_z$. The top and bottom black bars represent estimates of the complexity from the number of terms ((Right) the number of vertices per measurement group under a coloring scheme. In particular, while this number is constant for fermionic tomography, for the qubit operators this number increases at a nearly exponential rate.}
\label{fig:tomography}
\end{figure}

In the ACSE scheme, a clear advantage of the unencoded ACSE Ansatz is in the tomography of the ${}^2_Q A$ matrix, which as seen in prior work~\cite{Bonet-Monroig2019} results in a potentially logarithmic scaling entity. We show a basic comparison between the Jordan-Wigner transformation and the qubit-particle transformation in Fig.~\ref{fig:tomography}. The left side shows the effective scaling with respect to the number of qubits, i.e. $r$ in $O(q^r)$.  The right side shows the ratio of vertices to cliques in the corresponding graph problem, which is the ratio of the number of measured 2-RDM element contributions which can be recovered per grouping.

It is known that the grouping fermionic tomography is challenging due to the antisymmetry requirement that results in non-local groupings \cite{Bonet-Monroig2019,Izmaylov2019}. Despite this challenge, the $O(r^4)$ cost of 2-RDM tomography in molecular systems can be reduced by grouping to an $O(r^3)$ scheme~\cite{Gokhale2019} while even lower scaling schemes can be accomplished with additional swap circuits~\cite{Bonet-Monroig2019}, as well as with random unitary sampling techniques~\cite{Zhao2020}. We obtain our circuits through a graph theoretic approach with symmetry projection~\cite{Smart2021_tomo}. In the qubit case for a $k-$local operator, however, one can achieve a logarithmic scaling through known combinatorial schemes~\cite{Bonet-Monroig2019}. In our unencoded, qubit-particle case, a $k-$body excitation embodies only $k$- to $\frac{k}{2}$-body operators, and so does not span the full $k$-body operator space.  Consequently, the measurement scheme provides a super-linear scaling in required circuit preparation. As an example, for 28 qubits with over 200 million possible quantum states,  the 2-RDM has 92092 elements, and in the scheme represented in Fig.~\ref{fig:tomography} requires 6036 measurements in the encoded case, but only 92 measurements in the unencoded case.

Because the unencoded ACSE cannot be formulated in terms of qubit-RDMs alone, we implement $e^{i\delta H}$ through the tomography of an auxiliary state, which still requires the use of fermionic operators. While it might make sense to simply measure the partially encoded operators on a quantum computer, similar to measuring the 3-RDM in the encoded case, the scaling of the resulting 4-RDM like object is currently prohibitive for more than small systems. although utilizing measurement schemes, such as shadow tomography, might yield more efficient approximations in the future.


Table~\ref{tab:resource_count} presents the cost of evaluating the ACSE residual via Eq.~\ref{eq:auxiliarysolution} in both the encoded and unencoded CQE for a first-order Trotterization and a Cholesky decomposition of the Hamiltonian. The total number of CNOT counts is a product of the number of circuits and the scaling of the operator, in terms of the number of qubits. In the case of the Cholesky decomposition, we take the product of the number of distinct Cholesky terms and the number of unencoded or encoded 2-RDM terms. The unencoded ACSE with the Cholesky decomposition approach generally yields the most favorable scaling. Additionally, the average number of CNOT gates required for each auxiliary state using the Cholesky decomposition is orders of magnitude smaller than the number from the corresponding first-order trotterized approach, and hence, Cholesky decomposition is likely much more suitable for near-term approaches.

\begin{table}[]
    \caption{Upper bounds on the number of CNOT gates  for evaluation of the residuals of the unencoded and encoded ACSE on the quantum computer for numerous systems. The hydrogen chains (H$_2$ to H$_{12}$) are in a minimal basis, with the exception of H$_2$ (DZ), which has a double zeta basis set. \emph{Trotter} (or Trot.) refers to a first-order trotterization, \emph{CD $ \langle {\rm CNOT}\rangle $} to the average number of CNOT gates per decomposed term in the Cholesky decomposition, and \emph{Order} refers to the number of terms in the Cholesky decomposition. We use a threshold of $10^{-6}$ in the Cholesky decomposition, but more (or less) strict truncations can be taken. The bottom portion of the graph refers to the effective scaling in terms of total number of applied CNOT gates. }

    \centering
    \begin{tabular}{ccccccccccc}
    \hline\hline
          & & H$_2$ & H$_4$ & H$_6$ &  H$_8$ & H$_{10}$ & H$_{12}$ & N$_2$&  H$_2$ (DZ) & C$_2$H$_2$\\ \hline
        Trotter CNOT &  & 36 & 1088 & 12020 & 63760 & 191868 & 416168 & 38868 & 38868 & 102016 \\
        CD $ \langle {\rm CNOT} \rangle $& & 15 & 81 & 188 & 257 & 521 & 700 & 196 & 184 & 263 \\
        Order & & 4 & 8 & 12 & 16 & 20 & 24 &  49 & 55 & 70 \\ \hline
         \multicolumn{11}{c}{Total of CNOTs for Auxiliary Simulation, $\log_q$} \\ \hline
        Encoded, Trot.    & & 3.6 & 5.2 & 5.9 & 6.3 & 6.5 & 6.6 & 6.0 & 6.0 & 6.2 \\
        Encoded, CD    & & 3.9 & 4.9 & 5.3 & 5.3 & 5.6 & 5.6 & 5.5 & 5.5 & 5.6  \\
        Unencoded, Trot. & & 3.6 & 4.7 & 5.1 & 5.3 & 5.4 & 5.3 & 4.8 & 4.8 & 4.9 \\
        Unencoded, CD  & & 3.9 & 4.4 & 4.4 & 4.3 & 4.4 & 4.3 & 4.4 & 4.4 & 4.3 \\        \hline
    \end{tabular}
    \label{tab:resource_count}
\end{table}

\section{Discussion}

While the wave function from many electronic structure methods such as coupled cluster theory require significant alterations for implementation on quantum computers, the iterative solution of the ACSE naturally generates a wave function that is assembled from products of two-body unitary transformations that are amenable to implementation on a quantum computer.  Solution of the ACSE on a quantum computer---a CQE algorithm---does not require approximate reconstruction of the 3-RDM like its classical counterpart \cite{Mazziotti2006,Smart2021_prl}, and hence, at least in the absence of device noise, can yield energies and 2-RDMs that are in agreement with those from full configuration interaction.  In this paper we further develop the theory and results from \citet{Mazziotti2021} for solving the CQE with an unencoded wave function, a wave function expressed in terms of qubit particles rather than fermions.

Results show that the encoded and unencoded CQE yield similar accuracy across a variety of molecules at both equilibrium and non-equilibrium geometries.  As we discussed in Ref. \cite{Mazziotti2021} and the theory section here, the flexibility provided by the product of unencoded two-body operators is similar to the encoded operators. Unlike the hardware-efficient wave functions that do not specify the particle statistics and hence, have a many-to-one mapping to the fermion wave functions, the qubit-particle wave functions have an isomorphic mapping to the fermionic wave functions that can help to prevent optimization difficulties such as barren plateaus \cite{McClean2018}. Moreover, the highly commutative structure of the qubit-particle operators can be more easily leveraged than their equivalent fermionic operators to generate more compact state preparations and more efficient tomographies.

Qubit-particle wave functions have recently been applied in the coupled cluster singles-doubles (CCSD) Ansatz~\cite{Ryabinkin2018,Xia2020} as well as in hybrid VQE schemes like the ADAPT-VQE~\cite{Tang2019,Yordanov2021}. In UCCSD the use of qubit particles has been shown to produce a linear scaling reduction in the number of CNOT gates.  While the qubit-particle UCCSD can be less accurate than fermionic UCCSD due to the highly nonlinear nature of the Ansatz, in the calculations shown here the unencoded CQE can use its iterative formulation to continue its convergence towards the solution of the contracted equation.  The present theory and results provide a first step in exploring CQE algorithms for solving the unencoded ACSE.  Future work will consider further improvements from exploiting more compact wave functions as well as additional applications to larger molecules and materials.

\section{Conclusion}

Quantum simulation has the potential to reduce the cost of solving many-fermion problems.  Because quantum computers are based on qubits, however, their natural particles are not fermions but rather hard-core bosons known as qubit particles.  We have recently shown that there exists an isomorphism between fermion wave functions and qubit-particle wave functions, which suggests a natural parameterization of the {\em two-fermion} RDM in terms of the qubit-particle wave function that avoids fermionic encoding of the wave function.  Here we demonstrate that the recently proposed CQE algorithm for computing 2-RDMs by quantum simulation can be adapted to use unencoded qubit-particle wave functions rather than fermionic wave functions.  The unencoded CQE has similar theoretical accuracy as the encoded CQE, which can be converged to the exact, finite-basis solution of the Schr{\"o}dinger equation at least in the absence of quantum-device noise.  We illustrate the unencoded CQE's convergence, cost, and accuracy relative to that of the encoded CQE by quantum simulations of molecules at both equilibrium and non-equilibrium geometries.  Results show that the unencoded CQE has the potential in many cases to reduce the cost of quantum simulations of many-fermion problems without sacrificing accuracy even for strongly correlated systems.

\begin{acknowledgments}
D.A.M. gratefully acknowledges the Department of Energy, Office of Basic Energy Sciences, Grant DE-SC0019215 and the U.S. National Science Foundation Grants No. CHE-2035876, No. DMR-2037783, and  No. CHE-1565638.
\end{acknowledgments}
\appendix

\appendix

\section{ACSE Residuals from Hamiltonian Factorizations}

On near-term devices Hamiltonian simulation is challenging due to the generally high number of multi-qubit gates required, though numerous optimal approaches exist with varying resource requirements~\cite{Berry2015,Low2019,Lemieux2020}. Because we are interested in only a very small time step, we can exploit first-order approximations such as decomposing $\exp(i\delta \hat{H})$ as a sum of smaller, easier to implement operators, $\exp(i \delta H_p)$. If we take $|\Lambda^p_{m-1} \rangle$ to be auxiliary states of these operators, we can express the residuals as:
\begin{equation}
{}^2_Q A^{ij;kl}_m + \mathcal{O}(\delta^2) = \sum_p \delta^{-1} \Im \langle \Lambda_{m-1}^p| \qGam{i}{j}{l}{k} |\Lambda_{m-1}^p \rangle.
\end{equation}
The extreme of this strategy would be to simulate separately every term ${}^2 K^{pr}_{qs}$ of the Hamiltonian. This approach would not only lead to a substantial increase in the required sampling from the addition of $O(r^4)$ terms but also be most likely less efficient than exponentiating the 2-RDM operators and then taking the expectation of $\hat{H}$. Utilizing the commutative structure of the 2-RDM allows for more effective grouping, and a native $O(r^3)$ grouping pattern of Hamiltonian terms should be viable, similar to tomography-based grouping schemes.

Another approach involves decomposing the Hamiltonian, such as with the Cholesky decomposition of the two-electron integrals~\cite{Beebe1977, Motta2018, Hohenstein2010,Kivlichan2018}. While this offers benefits in both the encoded and unencoded 2-RDMs, this approach is potentially more practical in the latter case because of the substantial difference in the number of measurements required to measure the encoded and unencoded 2-RDMs, which we show in the Applications.

\section{Additional Computational Details}
A pivoted Cholesky decomposition \cite{HARBRECHT2012428} is utilized to obtain properly ordered terms in the Cholesky decomposition. The \textsc{python} module \textsc{hqca} as well as \textsc{qiskit} \cite{Smart_hqca__hybrid} are used, with electron integrals obtained through PySCF \cite{Qiskit, PySCF}.

\bibliography{
    ./biblio/0_heuristic_qc.bib,
    ./biblio/1_contracted.bib,
    ./biblio/2_qubit_encoding.bib,
    ./biblio/3_fermionic_simulation.bib,
    ./biblio/4_books.bib,
    ./biblio/5_hamiltonian_simulation.bib,
    ./biblio/6_tomography.bib,
    ./biblio/7_miscellaneous.bib
    }

\begin{thebibliography}{63}%
\makeatletter
\providecommand \@ifxundefined [1]{%
 \@ifx{#1\undefined}
}%
\providecommand \@ifnum [1]{%
 \ifnum #1\expandafter \@firstoftwo
 \else \expandafter \@secondoftwo
 \fi
}%
\providecommand \@ifx [1]{%
 \ifx #1\expandafter \@firstoftwo
 \else \expandafter \@secondoftwo
 \fi
}%
\providecommand \natexlab [1]{#1}%
\providecommand \enquote  [1]{``#1''}%
\providecommand \bibnamefont  [1]{#1}%
\providecommand \bibfnamefont [1]{#1}%
\providecommand \citenamefont [1]{#1}%
\providecommand \href@noop [0]{\@secondoftwo}%
\providecommand \href [0]{\begingroup \@sanitize@url \@href}%
\providecommand \@href[1]{\@@startlink{#1}\@@href}%
\providecommand \@@href[1]{\endgroup#1\@@endlink}%
\providecommand \@sanitize@url [0]{\catcode `\\12\catcode `\$12\catcode
  `\&12\catcode `\#12\catcode `\^12\catcode `\_12\catcode `\%12\relax}%
\providecommand \@@startlink[1]{}%
\providecommand \@@endlink[0]{}%
\providecommand \url  [0]{\begingroup\@sanitize@url \@url }%
\providecommand \@url [1]{\endgroup\@href {#1}{\urlprefix }}%
\providecommand \urlprefix  [0]{URL }%
\providecommand \Eprint [0]{\href }%
\providecommand \doibase [0]{https://doi.org/}%
\providecommand \selectlanguage [0]{\@gobble}%
\providecommand \bibinfo  [0]{\@secondoftwo}%
\providecommand \bibfield  [0]{\@secondoftwo}%
\providecommand \translation [1]{[#1]}%
\providecommand \BibitemOpen [0]{}%
\providecommand \bibitemStop [0]{}%
\providecommand \bibitemNoStop [0]{.\EOS\space}%
\providecommand \EOS [0]{\spacefactor3000\relax}%
\providecommand \BibitemShut  [1]{\csname bibitem#1\endcsname}%
\let\auto@bib@innerbib\@empty
\bibitem [{\citenamefont {Abrams}\ and\ \citenamefont
  {Lloyd}(1997)}]{Abrams1997}%
  \BibitemOpen
  \bibfield  {author} {\bibinfo {author} {\bibfnamefont {D.~S.}\ \bibnamefont
  {Abrams}}\ and\ \bibinfo {author} {\bibfnamefont {S.}~\bibnamefont {Lloyd}},\
  }\bibfield  {title} {\bibinfo {title} {{Simulation of Many-Body Fermi Systems
  on a Universal Quantum Computer}},\ }\href
  {https://doi.org/10.1103/PhysRevLett.79.2586} {\bibfield  {journal} {\bibinfo
   {journal} {Physical Review Letters}\ }\textbf {\bibinfo {volume} {79}},\
  \bibinfo {pages} {2586} (\bibinfo {year} {1997})},\ \Eprint
  {https://arxiv.org/abs/9703054} {arXiv:9703054 [quant-ph]} \BibitemShut
  {NoStop}%
\bibitem [{\citenamefont {Whitfield}\ \emph {et~al.}(2011)\citenamefont
  {Whitfield}, \citenamefont {Biamonte},\ and\ \citenamefont
  {Aspuru-Guzik}}]{Whitfield2011}%
  \BibitemOpen
  \bibfield  {author} {\bibinfo {author} {\bibfnamefont {J.~D.}\ \bibnamefont
  {Whitfield}}, \bibinfo {author} {\bibfnamefont {J.}~\bibnamefont
  {Biamonte}},\ and\ \bibinfo {author} {\bibfnamefont {A.}~\bibnamefont
  {Aspuru-Guzik}},\ }\bibfield  {title} {\bibinfo {title} {{Simulation of
  electronic structure Hamiltonians using quantum computers}},\ }\href
  {https://doi.org/10.1080/00268976.2011.552441} {\bibfield  {journal}
  {\bibinfo  {journal} {Molecular Physics}\ }\textbf {\bibinfo {volume}
  {109}},\ \bibinfo {pages} {735} (\bibinfo {year} {2011})},\ \Eprint
  {https://arxiv.org/abs/1001.3855} {arXiv:1001.3855} \BibitemShut {NoStop}%
\bibitem [{\citenamefont {Mazziotti}\ \emph {et~al.}(2021)\citenamefont
  {Mazziotti}, \citenamefont {Smart},\ and\ \citenamefont
  {Mazziotti}}]{Mazziotti2021}%
  \BibitemOpen
  \bibfield  {author} {\bibinfo {author} {\bibfnamefont {D.~A.}\ \bibnamefont
  {Mazziotti}}, \bibinfo {author} {\bibfnamefont {S.~E.}\ \bibnamefont
  {Smart}},\ and\ \bibinfo {author} {\bibfnamefont {A.~R.}\ \bibnamefont
  {Mazziotti}},\ }\bibfield  {title} {\bibinfo {title} {{Quantum simulation of
  molecules without fermionic encoding of the wave function}},\ }\href
  {https://doi.org/10.1088/1367-2630/ac3573} {\bibfield  {journal} {\bibinfo
  {journal} {New Journal of Physics}\ }\textbf {\bibinfo {volume} {23}},\
  \bibinfo {pages} {113037} (\bibinfo {year} {2021})},\ \Eprint
  {https://arxiv.org/abs/2101.11607} {arXiv:2101.11607} \BibitemShut {NoStop}%
\bibitem [{\citenamefont {Wu}\ and\ \citenamefont {Lidar}(2002)}]{Wu2002}%
  \BibitemOpen
  \bibfield  {author} {\bibinfo {author} {\bibfnamefont {L.-A.}\ \bibnamefont
  {Wu}}\ and\ \bibinfo {author} {\bibfnamefont {D.~A.}\ \bibnamefont {Lidar}},\
  }\bibfield  {title} {\bibinfo {title} {{Qubits as parafermions}},\ }\href
  {https://doi.org/10.1063/1.1499208} {\bibfield  {journal} {\bibinfo
  {journal} {Journal of Mathematical Physics}\ }\textbf {\bibinfo {volume}
  {43}},\ \bibinfo {pages} {4506} (\bibinfo {year} {2002})},\ \Eprint
  {https://arxiv.org/abs/0109078} {arXiv:0109078 [quant-ph]} \BibitemShut
  {NoStop}%
\bibitem [{\citenamefont {Feynman}(1982)}]{Feynman1982}%
  \BibitemOpen
  \bibfield  {author} {\bibinfo {author} {\bibfnamefont {R.~P.}\ \bibnamefont
  {Feynman}},\ }\bibfield  {title} {\bibinfo {title} {{Simulating physics with
  computers}},\ }\href {https://doi.org/10.1007/BF02650179} {\bibfield
  {journal} {\bibinfo  {journal} {International Journal of Theoretical
  Physics}\ }\textbf {\bibinfo {volume} {21}},\ \bibinfo {pages} {467}
  (\bibinfo {year} {1982})}\BibitemShut {NoStop}%
\bibitem [{\citenamefont {Jordan}\ and\ \citenamefont
  {Wigner}(1928)}]{Jordan1928}%
  \BibitemOpen
  \bibfield  {author} {\bibinfo {author} {\bibfnamefont {P.}~\bibnamefont
  {Jordan}}\ and\ \bibinfo {author} {\bibfnamefont {E.}~\bibnamefont
  {Wigner}},\ }\bibfield  {title} {\bibinfo {title} {{{\"U}ber das Paulische
  {\"A}quivalenzverbot}},\ }\href {https://doi.org/10.1007/BF01331938}
  {\bibfield  {journal} {\bibinfo  {journal} {Zeitschrift f{\"u}r Physik}\
  }\textbf {\bibinfo {volume} {47}},\ \bibinfo {pages} {631} (\bibinfo {year}
  {1928})}\BibitemShut {NoStop}%
\bibitem [{\citenamefont {Bravyi}\ and\ \citenamefont
  {Kitaev}(2002)}]{Bravyi2000}%
  \BibitemOpen
  \bibfield  {author} {\bibinfo {author} {\bibfnamefont {S.~B.}\ \bibnamefont
  {Bravyi}}\ and\ \bibinfo {author} {\bibfnamefont {A.~Y.}\ \bibnamefont
  {Kitaev}},\ }\bibfield  {title} {\bibinfo {title} {{Fermionic Quantum
  Computation}},\ }\href {https://doi.org/10.1006/aphy.2002.6254} {\bibfield
  {journal} {\bibinfo  {journal} {Annals of Physics}\ }\textbf {\bibinfo
  {volume} {298}},\ \bibinfo {pages} {210} (\bibinfo {year} {2002})},\ \Eprint
  {https://arxiv.org/abs/0003137} {arXiv:0003137 [quant-ph]} \BibitemShut
  {NoStop}%
\bibitem [{\citenamefont {Seeley}\ \emph {et~al.}(2012)\citenamefont {Seeley},
  \citenamefont {Richard},\ and\ \citenamefont {Love}}]{Seeley2012}%
  \BibitemOpen
  \bibfield  {author} {\bibinfo {author} {\bibfnamefont {J.~T.}\ \bibnamefont
  {Seeley}}, \bibinfo {author} {\bibfnamefont {M.~J.}\ \bibnamefont
  {Richard}},\ and\ \bibinfo {author} {\bibfnamefont {P.~J.}\ \bibnamefont
  {Love}},\ }\bibfield  {title} {\bibinfo {title} {{The Bravyi-Kitaev
  transformation for quantum computation of electronic structure}},\ }\href
  {https://doi.org/10.1063/1.4768229} {\bibfield  {journal} {\bibinfo
  {journal} {The Journal of Chemical Physics}\ }\textbf {\bibinfo {volume}
  {137}},\ \bibinfo {pages} {224109} (\bibinfo {year} {2012})},\ \Eprint
  {https://arxiv.org/abs/1208.5986} {arXiv:1208.5986} \BibitemShut {NoStop}%
\bibitem [{\citenamefont {Hastings}\ \emph {et~al.}(2014)\citenamefont
  {Hastings}, \citenamefont {Wecker}, \citenamefont {Bauer},\ and\
  \citenamefont {Troyer}}]{Hastings2014}%
  \BibitemOpen
  \bibfield  {author} {\bibinfo {author} {\bibfnamefont {M.~B.}\ \bibnamefont
  {Hastings}}, \bibinfo {author} {\bibfnamefont {D.}~\bibnamefont {Wecker}},
  \bibinfo {author} {\bibfnamefont {B.}~\bibnamefont {Bauer}},\ and\ \bibinfo
  {author} {\bibfnamefont {M.}~\bibnamefont {Troyer}},\ }\bibfield  {title}
  {\bibinfo {title} {{Improving Quantum Algorithms for Quantum Chemistry}},\
  }\href {http://arxiv.org/abs/1403.1539} {\bibfield  {journal} {\bibinfo
  {journal} {arXiv}\ } (\bibinfo {year} {2014})},\ \Eprint
  {https://arxiv.org/abs/1403.1539} {arXiv:1403.1539} \BibitemShut {NoStop}%
\bibitem [{\citenamefont {Wecker}\ \emph {et~al.}(2014)\citenamefont {Wecker},
  \citenamefont {Bauer}, \citenamefont {Clark}, \citenamefont {Hastings},\ and\
  \citenamefont {Troyer}}]{Wecker2014}%
  \BibitemOpen
  \bibfield  {author} {\bibinfo {author} {\bibfnamefont {D.}~\bibnamefont
  {Wecker}}, \bibinfo {author} {\bibfnamefont {B.}~\bibnamefont {Bauer}},
  \bibinfo {author} {\bibfnamefont {B.~K.}\ \bibnamefont {Clark}}, \bibinfo
  {author} {\bibfnamefont {M.~B.}\ \bibnamefont {Hastings}},\ and\ \bibinfo
  {author} {\bibfnamefont {M.}~\bibnamefont {Troyer}},\ }\bibfield  {title}
  {\bibinfo {title} {{Gate-count estimates for performing quantum chemistry on
  small quantum computers}},\ }\href
  {https://doi.org/10.1103/PhysRevA.90.022305} {\bibfield  {journal} {\bibinfo
  {journal} {Physical Review A - Atomic, Molecular, and Optical Physics}\
  }\textbf {\bibinfo {volume} {90}},\ \bibinfo {pages} {1} (\bibinfo {year}
  {2014})},\ \Eprint {https://arxiv.org/abs/1312.1695} {arXiv:1312.1695}
  \BibitemShut {NoStop}%
\bibitem [{\citenamefont {Kandala}\ \emph {et~al.}(2017)\citenamefont
  {Kandala}, \citenamefont {Mezzacapo}, \citenamefont {Temme}, \citenamefont
  {Takita}, \citenamefont {Brink}, \citenamefont {Chow},\ and\ \citenamefont
  {Gambetta}}]{Kandala2017}%
  \BibitemOpen
  \bibfield  {author} {\bibinfo {author} {\bibfnamefont {A.}~\bibnamefont
  {Kandala}}, \bibinfo {author} {\bibfnamefont {A.}~\bibnamefont {Mezzacapo}},
  \bibinfo {author} {\bibfnamefont {K.}~\bibnamefont {Temme}}, \bibinfo
  {author} {\bibfnamefont {M.}~\bibnamefont {Takita}}, \bibinfo {author}
  {\bibfnamefont {M.}~\bibnamefont {Brink}}, \bibinfo {author} {\bibfnamefont
  {J.~M.}\ \bibnamefont {Chow}},\ and\ \bibinfo {author} {\bibfnamefont
  {J.~M.}\ \bibnamefont {Gambetta}},\ }\bibfield  {title} {\bibinfo {title}
  {{Hardware-efficient variational quantum eigensolver for small molecules and
  quantum magnets}},\ }\href {https://doi.org/10.1038/nature23879} {\bibfield
  {journal} {\bibinfo  {journal} {Nature}\ }\textbf {\bibinfo {volume} {549}},\
  \bibinfo {pages} {242} (\bibinfo {year} {2017})},\ \Eprint
  {https://arxiv.org/abs/1704.05018} {arXiv:1704.05018} \BibitemShut {NoStop}%
\bibitem [{\citenamefont {Choquette}\ \emph {et~al.}(2020)\citenamefont
  {Choquette}, \citenamefont {{Di Paolo}}, \citenamefont {Barkoutsos},
  \citenamefont {S{\'{e}}n{\'{e}}chal}, \citenamefont {Tavernelli},\ and\
  \citenamefont {Blais}}]{Choquette2020}%
  \BibitemOpen
  \bibfield  {author} {\bibinfo {author} {\bibfnamefont {A.}~\bibnamefont
  {Choquette}}, \bibinfo {author} {\bibfnamefont {A.}~\bibnamefont {{Di
  Paolo}}}, \bibinfo {author} {\bibfnamefont {P.~K.}\ \bibnamefont
  {Barkoutsos}}, \bibinfo {author} {\bibfnamefont {D.}~\bibnamefont
  {S{\'{e}}n{\'{e}}chal}}, \bibinfo {author} {\bibfnamefont {I.}~\bibnamefont
  {Tavernelli}},\ and\ \bibinfo {author} {\bibfnamefont {A.}~\bibnamefont
  {Blais}},\ }\bibfield  {title} {\bibinfo {title}
  {{Quantum-optimal-control-inspired ansatz for variational quantum
  algorithms}},\ }\href@noop {} {\bibfield  {journal} {\bibinfo  {journal}
  {arXiv}\ ,\ \bibinfo {pages} {1}} (\bibinfo {year} {2020})},\ \Eprint
  {https://arxiv.org/abs/2008.01098} {arXiv:2008.01098} \BibitemShut {NoStop}%
\bibitem [{\citenamefont {Barron}\ \emph {et~al.}(2020)\citenamefont {Barron},
  \citenamefont {Gard}, \citenamefont {Altman}, \citenamefont {Mayhall},
  \citenamefont {Barnes}, \citenamefont {Economou},\ and\ \citenamefont
  {Tech}}]{Barron2020}%
  \BibitemOpen
  \bibfield  {author} {\bibinfo {author} {\bibfnamefont {G.~S.}\ \bibnamefont
  {Barron}}, \bibinfo {author} {\bibfnamefont {B.~T.}\ \bibnamefont {Gard}},
  \bibinfo {author} {\bibfnamefont {O.~J.}\ \bibnamefont {Altman}}, \bibinfo
  {author} {\bibfnamefont {N.~J.}\ \bibnamefont {Mayhall}}, \bibinfo {author}
  {\bibfnamefont {E.}~\bibnamefont {Barnes}}, \bibinfo {author} {\bibfnamefont
  {S.~E.}\ \bibnamefont {Economou}},\ and\ \bibinfo {author} {\bibfnamefont
  {V.}~\bibnamefont {Tech}},\ }\bibfield  {title} {\bibinfo {title}
  {{Preserving Symmetries for Variational Quantum Eigensolvers in the Presence
  of Noise}},\ }\href@noop {} {\bibfield  {journal} {\bibinfo  {journal}
  {arXiv}\ ,\ \bibinfo {pages} {1}} (\bibinfo {year} {2020})},\ \Eprint
  {https://arxiv.org/abs/arXiv:2003.00171v1} {arXiv:arXiv:2003.00171v1}
  \BibitemShut {NoStop}%
\bibitem [{\citenamefont {Xia}\ and\ \citenamefont {Kais}(2020)}]{Xia2020}%
  \BibitemOpen
  \bibfield  {author} {\bibinfo {author} {\bibfnamefont {R.}~\bibnamefont
  {Xia}}\ and\ \bibinfo {author} {\bibfnamefont {S.}~\bibnamefont {Kais}},\
  }\bibfield  {title} {\bibinfo {title} {{Qubit coupled cluster singles and
  doubles variational quantum eigensolver ansatz for electronic structure
  calculations}},\ }\href {https://doi.org/10.1088/2058-9565/abbc74} {\bibfield
   {journal} {\bibinfo  {journal} {Quantum Science and Technology}\ }\textbf
  {\bibinfo {volume} {6}},\ \bibinfo {pages} {15001} (\bibinfo {year}
  {2020})},\ \Eprint {https://arxiv.org/abs/2005.08451} {arXiv:2005.08451}
  \BibitemShut {NoStop}%
\bibitem [{\citenamefont {Izmaylov}\ \emph {et~al.}(2020)\citenamefont
  {Izmaylov}, \citenamefont {Yen}, \citenamefont {Lang},\ and\ \citenamefont
  {Verteletskyi}}]{Izmaylov2019}%
  \BibitemOpen
  \bibfield  {author} {\bibinfo {author} {\bibfnamefont {A.~F.}\ \bibnamefont
  {Izmaylov}}, \bibinfo {author} {\bibfnamefont {T.-C.}\ \bibnamefont {Yen}},
  \bibinfo {author} {\bibfnamefont {R.~A.}\ \bibnamefont {Lang}},\ and\
  \bibinfo {author} {\bibfnamefont {V.}~\bibnamefont {Verteletskyi}},\
  }\bibfield  {title} {\bibinfo {title} {{Unitary Partitioning Approach to the
  Measurement Problem in the Variational Quantum Eigensolver Method}},\ }\href
  {https://doi.org/10.1021/acs.jctc.9b00791} {\bibfield  {journal} {\bibinfo
  {journal} {Journal of Chemical Theory and Computation}\ }\textbf {\bibinfo
  {volume} {16}},\ \bibinfo {pages} {190} (\bibinfo {year} {2020})},\ \Eprint
  {https://arxiv.org/abs/1907.09040} {arXiv:1907.09040} \BibitemShut {NoStop}%
\bibitem [{\citenamefont {Ryabinkin}\ \emph {et~al.}(2018)\citenamefont
  {Ryabinkin}, \citenamefont {Yen}, \citenamefont {Genin},\ and\ \citenamefont
  {Izmaylov}}]{Ryabinkin2018}%
  \BibitemOpen
  \bibfield  {author} {\bibinfo {author} {\bibfnamefont {I.~G.}\ \bibnamefont
  {Ryabinkin}}, \bibinfo {author} {\bibfnamefont {T.~C.}\ \bibnamefont {Yen}},
  \bibinfo {author} {\bibfnamefont {S.~N.}\ \bibnamefont {Genin}},\ and\
  \bibinfo {author} {\bibfnamefont {A.~F.}\ \bibnamefont {Izmaylov}},\
  }\bibfield  {title} {\bibinfo {title} {{Qubit Coupled Cluster Method: A
  Systematic Approach to Quantum Chemistry on a Quantum Computer}},\ }\href
  {https://doi.org/10.1021/acs.jctc.8b00932} {\bibfield  {journal} {\bibinfo
  {journal} {Journal of Chemical Theory and Computation}\ }\textbf {\bibinfo
  {volume} {14}},\ \bibinfo {pages} {6317} (\bibinfo {year} {2018})},\ \Eprint
  {https://arxiv.org/abs/1809.03827} {arXiv:1809.03827} \BibitemShut {NoStop}%
\bibitem [{\citenamefont {Ryabinkin}\ \emph {et~al.}(2020)\citenamefont
  {Ryabinkin}, \citenamefont {Lang}, \citenamefont {Genin},\ and\ \citenamefont
  {Izmaylov}}]{Ryabinkin2020}%
  \BibitemOpen
  \bibfield  {author} {\bibinfo {author} {\bibfnamefont {I.~G.}\ \bibnamefont
  {Ryabinkin}}, \bibinfo {author} {\bibfnamefont {R.~A.}\ \bibnamefont {Lang}},
  \bibinfo {author} {\bibfnamefont {S.~N.}\ \bibnamefont {Genin}},\ and\
  \bibinfo {author} {\bibfnamefont {A.~F.}\ \bibnamefont {Izmaylov}},\
  }\bibfield  {title} {\bibinfo {title} {{Iterative Qubit Coupled Cluster
  Approach with Efficient Screening of Generators}},\ }\href
  {https://doi.org/10.1021/acs.jctc.9b01084} {\bibfield  {journal} {\bibinfo
  {journal} {Journal of Chemical Theory and Computation}\ }\textbf {\bibinfo
  {volume} {16}},\ \bibinfo {pages} {1055} (\bibinfo {year} {2020})},\ \Eprint
  {https://arxiv.org/abs/1906.11192} {arXiv:1906.11192} \BibitemShut {NoStop}%
\bibitem [{\citenamefont {Ryabinkin}\ \emph {et~al.}(2021)\citenamefont
  {Ryabinkin}, \citenamefont {Izmaylov},\ and\ \citenamefont
  {Genin}}]{Ryabinkin2021}%
  \BibitemOpen
  \bibfield  {author} {\bibinfo {author} {\bibfnamefont {I.~G.}\ \bibnamefont
  {Ryabinkin}}, \bibinfo {author} {\bibfnamefont {A.~F.}\ \bibnamefont
  {Izmaylov}},\ and\ \bibinfo {author} {\bibfnamefont {S.~N.}\ \bibnamefont
  {Genin}},\ }\bibfield  {title} {\bibinfo {title} {{A posteriori corrections
  to the iterative qubit coupled cluster method to minimize the use of quantum
  resources in large-scale calculations}},\ }\bibfield  {journal} {\bibinfo
  {journal} {Quantum Science and Technology}\ }\textbf {\bibinfo {volume}
  {6}},\ \href {https://doi.org/10.1088/2058-9565/abda8e}
  {10.1088/2058-9565/abda8e} (\bibinfo {year} {2021}),\ \Eprint
  {https://arxiv.org/abs/2009.13622} {arXiv:2009.13622} \BibitemShut {NoStop}%
\bibitem [{\citenamefont {Tang}\ \emph {et~al.}(2021)\citenamefont {Tang},
  \citenamefont {Shkolnikov}, \citenamefont {Barron}, \citenamefont {Grimsley},
  \citenamefont {Mayhall}, \citenamefont {Barnes},\ and\ \citenamefont
  {Economou}}]{Tang2019}%
  \BibitemOpen
  \bibfield  {author} {\bibinfo {author} {\bibfnamefont {H.~L.}\ \bibnamefont
  {Tang}}, \bibinfo {author} {\bibfnamefont {V.}~\bibnamefont {Shkolnikov}},
  \bibinfo {author} {\bibfnamefont {G.~S.}\ \bibnamefont {Barron}}, \bibinfo
  {author} {\bibfnamefont {H.~R.}\ \bibnamefont {Grimsley}}, \bibinfo {author}
  {\bibfnamefont {N.~J.}\ \bibnamefont {Mayhall}}, \bibinfo {author}
  {\bibfnamefont {E.}~\bibnamefont {Barnes}},\ and\ \bibinfo {author}
  {\bibfnamefont {S.~E.}\ \bibnamefont {Economou}},\ }\bibfield  {title}
  {\bibinfo {title} {{Qubit-ADAPT-VQE: An Adaptive Algorithm for Constructing
  Hardware-Efficient Ans{\"{a}}tze on a Quantum Processor}},\ }\href
  {https://doi.org/10.1103/PRXQuantum.2.020310} {\bibfield  {journal} {\bibinfo
   {journal} {PRX Quantum}\ }\textbf {\bibinfo {volume} {2}},\ \bibinfo {pages}
  {020310} (\bibinfo {year} {2021})},\ \Eprint
  {https://arxiv.org/abs/1911.10205} {arXiv:1911.10205} \BibitemShut {NoStop}%
\bibitem [{\citenamefont {Smart}\ and\ \citenamefont
  {Mazziotti}(2021{\natexlab{a}})}]{Smart2021_prl}%
  \BibitemOpen
  \bibfield  {author} {\bibinfo {author} {\bibfnamefont {S.~E.}\ \bibnamefont
  {Smart}}\ and\ \bibinfo {author} {\bibfnamefont {D.~A.}\ \bibnamefont
  {Mazziotti}},\ }\bibfield  {title} {\bibinfo {title} {{Quantum Solver of
  Contracted Eigenvalue Equations for Scalable Molecular Simulations on Quantum
  Computing Devices}},\ }\href {https://doi.org/10.1103/PhysRevLett.126.070504}
  {\bibfield  {journal} {\bibinfo  {journal} {Physical Review Letters}\
  }\textbf {\bibinfo {volume} {126}},\ \bibinfo {pages} {070504} (\bibinfo
  {year} {2021}{\natexlab{a}})},\ \Eprint {https://arxiv.org/abs/2004.11416}
  {arXiv:2004.11416} \BibitemShut {NoStop}%
\bibitem [{\citenamefont {Boyn}\ \emph {et~al.}(2021)\citenamefont {Boyn},
  \citenamefont {Lykhin}, \citenamefont {Smart}, \citenamefont {Gagliardi},\
  and\ \citenamefont {Mazziotti}}]{Boyn2021}%
  \BibitemOpen
  \bibfield  {author} {\bibinfo {author} {\bibfnamefont {J.-N.}\ \bibnamefont
  {Boyn}}, \bibinfo {author} {\bibfnamefont {A.~O.}\ \bibnamefont {Lykhin}},
  \bibinfo {author} {\bibfnamefont {S.~E.}\ \bibnamefont {Smart}}, \bibinfo
  {author} {\bibfnamefont {L.}~\bibnamefont {Gagliardi}},\ and\ \bibinfo
  {author} {\bibfnamefont {D.~A.}\ \bibnamefont {Mazziotti}},\ }\bibfield
  {title} {\bibinfo {title} {{Quantum-classical hybrid algorithm for the
  simulation of all-electron correlation}},\ }\href
  {https://doi.org/10.1063/5.0074842} {\bibfield  {journal} {\bibinfo
  {journal} {The Journal of Chemical Physics}\ }\textbf {\bibinfo {volume}
  {155}},\ \bibinfo {pages} {244106} (\bibinfo {year} {2021})},\ \Eprint
  {https://arxiv.org/abs/2106.11972} {arXiv:2106.11972} \BibitemShut {NoStop}%
\bibitem [{\citenamefont {Smart}\ \emph {et~al.}(2022)\citenamefont {Smart},
  \citenamefont {Boyn},\ and\ \citenamefont {Mazziotti}}]{Smart2022_benzyne}%
  \BibitemOpen
  \bibfield  {author} {\bibinfo {author} {\bibfnamefont {S.~E.}\ \bibnamefont
  {Smart}}, \bibinfo {author} {\bibfnamefont {J.-N.}\ \bibnamefont {Boyn}},\
  and\ \bibinfo {author} {\bibfnamefont {D.~A.}\ \bibnamefont {Mazziotti}},\
  }\bibfield  {title} {\bibinfo {title} {{Resolving correlated states of
  benzyne with an error-mitigated contracted quantum eigensolver}},\ }\href
  {https://doi.org/10.1103/PhysRevA.105.022405} {\bibfield  {journal} {\bibinfo
   {journal} {Physical Review A}\ }\textbf {\bibinfo {volume} {105}},\ \bibinfo
  {pages} {022405} (\bibinfo {year} {2022})},\ \Eprint
  {https://arxiv.org/abs/2103.06876} {arXiv:2103.06876} \BibitemShut {NoStop}%
\bibitem [{\citenamefont {Mazziotti}(2006)}]{Mazziotti2006}%
  \BibitemOpen
  \bibfield  {author} {\bibinfo {author} {\bibfnamefont {D.~A.}\ \bibnamefont
  {Mazziotti}},\ }\bibfield  {title} {\bibinfo {title} {{Anti-Hermitian
  Contracted Schr{\"{o}}dinger Equation: Direct Determination of the
  Two-Electron Reduced Density Matrices of Many-Electron Molecules}},\ }\href
  {https://doi.org/10.1103/PhysRevLett.97.143002} {\bibfield  {journal}
  {\bibinfo  {journal} {Physical Review Letters}\ }\textbf {\bibinfo {volume}
  {97}},\ \bibinfo {pages} {143002} (\bibinfo {year} {2006})}\BibitemShut
  {NoStop}%
\bibitem [{\citenamefont {Mazziotti}(2007{\natexlab{a}})}]{Mazziotti2007}%
  \BibitemOpen
  \bibfield  {author} {\bibinfo {author} {\bibfnamefont {D.~A.}\ \bibnamefont
  {Mazziotti}},\ }\bibfield  {title} {\bibinfo {title} {{Anti-Hermitian part of
  the contracted Schr{\"{o}}dinger equation for the direct calculation of
  two-electron reduced density matrices}},\ }\href
  {https://doi.org/10.1103/PhysRevA.75.022505} {\bibfield  {journal} {\bibinfo
  {journal} {Physical Review A - Atomic, Molecular, and Optical Physics}\
  }\textbf {\bibinfo {volume} {75}},\ \bibinfo {pages} {1} (\bibinfo {year}
  {2007}{\natexlab{a}})}\BibitemShut {NoStop}%
\bibitem [{\citenamefont {Mazziotti}(2007{\natexlab{b}})}]{Mazziotti2007_mr}%
  \BibitemOpen
  \bibfield  {author} {\bibinfo {author} {\bibfnamefont {D.~A.}\ \bibnamefont
  {Mazziotti}},\ }\bibfield  {title} {\bibinfo {title} {{Multireference
  many-electron correlation energies from two-electron reduced density matrices
  computed by solving the anti-Hermitian contracted Schr??dinger equation}},\
  }\href {https://doi.org/10.1103/PhysRevA.76.052502} {\bibfield  {journal}
  {\bibinfo  {journal} {Physical Review A - Atomic, Molecular, and Optical
  Physics}\ }\textbf {\bibinfo {volume} {76}},\ \bibinfo {pages} {1} (\bibinfo
  {year} {2007}{\natexlab{b}})}\BibitemShut {NoStop}%
\bibitem [{\citenamefont {Gidofalvi}\ and\ \citenamefont
  {Mazziotti}(2007)}]{Gidofalvi2007}%
  \BibitemOpen
  \bibfield  {author} {\bibinfo {author} {\bibfnamefont {G.}~\bibnamefont
  {Gidofalvi}}\ and\ \bibinfo {author} {\bibfnamefont {D.~A.}\ \bibnamefont
  {Mazziotti}},\ }\bibfield  {title} {\bibinfo {title} {{Multireference
  self-consistent-field energies without the many-electron wave function
  through a variational low-rank two-electron reduced-density-matrix method}},\
  }\bibfield  {journal} {\bibinfo  {journal} {Journal of Chemical Physics}\
  }\textbf {\bibinfo {volume} {127}},\ \href
  {https://doi.org/10.1063/1.2817602} {10.1063/1.2817602} (\bibinfo {year}
  {2007})\BibitemShut {NoStop}%
\bibitem [{\citenamefont {Rothman}\ \emph {et~al.}(2009)\citenamefont
  {Rothman}, \citenamefont {Foley},\ and\ \citenamefont
  {Mazziotti}}]{Rothman2009}%
  \BibitemOpen
  \bibfield  {author} {\bibinfo {author} {\bibfnamefont {A.~E.}\ \bibnamefont
  {Rothman}}, \bibinfo {author} {\bibfnamefont {J.~J.}\ \bibnamefont {Foley}},\
  and\ \bibinfo {author} {\bibfnamefont {D.~A.}\ \bibnamefont {Mazziotti}},\
  }\bibfield  {title} {\bibinfo {title} {{Open-shell energies and two-electron
  reduced density matrices from the anti-Hermitian contracted Schr{\"{o}}dinger
  equation: A spin-coupled approach}},\ }\href
  {https://doi.org/10.1103/physreva.80.052508} {\bibfield  {journal} {\bibinfo
  {journal} {Physical Review A}\ }\textbf {\bibinfo {volume} {80}},\ \bibinfo
  {pages} {1} (\bibinfo {year} {2009})}\BibitemShut {NoStop}%
\bibitem [{\citenamefont {Snyder}\ and\ \citenamefont
  {Mazziotti}(2012)}]{Snyder2012}%
  \BibitemOpen
  \bibfield  {author} {\bibinfo {author} {\bibfnamefont {J.~W.}\ \bibnamefont
  {Snyder}}\ and\ \bibinfo {author} {\bibfnamefont {D.~A.}\ \bibnamefont
  {Mazziotti}},\ }\bibfield  {title} {\bibinfo {title} {{Photoexcited
  tautomerization of vinyl alcohol to acetylaldehydevia a conical intersection
  from contracted Schr{\"{o}}dinger theory}},\ }\href
  {https://doi.org/10.1039/C2CP23065H} {\bibfield  {journal} {\bibinfo
  {journal} {Phys. Chem. Chem. Phys.}\ }\textbf {\bibinfo {volume} {14}},\
  \bibinfo {pages} {1660} (\bibinfo {year} {2012})}\BibitemShut {NoStop}%
\bibitem [{\citenamefont {Sand}\ and\ \citenamefont
  {Mazziotti}(2015)}]{Sand2015}%
  \BibitemOpen
  \bibfield  {author} {\bibinfo {author} {\bibfnamefont {A.~M.}\ \bibnamefont
  {Sand}}\ and\ \bibinfo {author} {\bibfnamefont {D.~A.}\ \bibnamefont
  {Mazziotti}},\ }\bibfield  {title} {\bibinfo {title} {{Enhanced computational
  efficiency in the direct determination of the two-electron reduced density
  matrix from the anti-Hermitian contracted Schr{\"{o}}dinger equation with
  application to ground and excited states of conjugated $\pi$-systems}},\
  }\href {https://doi.org/10.1063/1.4931471} {\bibfield  {journal} {\bibinfo
  {journal} {The Journal of Chemical Physics}\ }\textbf {\bibinfo {volume}
  {143}},\ \bibinfo {pages} {134110} (\bibinfo {year} {2015})}\BibitemShut
  {NoStop}%
\bibitem [{\citenamefont {Boyn}\ and\ \citenamefont
  {Mazziotti}(2021)}]{Boyn2021_spin}%
  \BibitemOpen
  \bibfield  {author} {\bibinfo {author} {\bibfnamefont {J.~N.}\ \bibnamefont
  {Boyn}}\ and\ \bibinfo {author} {\bibfnamefont {D.~A.}\ \bibnamefont
  {Mazziotti}},\ }\bibfield  {title} {\bibinfo {title} {{Accurate
  singlet-triplet gaps in biradicals via the spin averaged anti-Hermitian
  contracted Schr{\"{o}}dinger equation}},\ }\bibfield  {journal} {\bibinfo
  {journal} {Journal of Chemical Physics}\ }\textbf {\bibinfo {volume} {154}},\
  \href {https://doi.org/10.1063/5.0045007} {10.1063/5.0045007} (\bibinfo
  {year} {2021}),\ \Eprint {https://arxiv.org/abs/2104.00626}
  {arXiv:2104.00626} \BibitemShut {NoStop}%
\bibitem [{\citenamefont {Peruzzo}\ \emph {et~al.}(2014)\citenamefont
  {Peruzzo}, \citenamefont {McClean}, \citenamefont {Shadbolt}, \citenamefont
  {Yung}, \citenamefont {Zhou}, \citenamefont {Love}, \citenamefont
  {Aspuru-Guzik},\ and\ \citenamefont {O'Brien}}]{Peruzzo2014}%
  \BibitemOpen
  \bibfield  {author} {\bibinfo {author} {\bibfnamefont {A.}~\bibnamefont
  {Peruzzo}}, \bibinfo {author} {\bibfnamefont {J.}~\bibnamefont {McClean}},
  \bibinfo {author} {\bibfnamefont {P.}~\bibnamefont {Shadbolt}}, \bibinfo
  {author} {\bibfnamefont {M.-H.}\ \bibnamefont {Yung}}, \bibinfo {author}
  {\bibfnamefont {X.-Q.}\ \bibnamefont {Zhou}}, \bibinfo {author}
  {\bibfnamefont {P.~J.}\ \bibnamefont {Love}}, \bibinfo {author}
  {\bibfnamefont {A.}~\bibnamefont {Aspuru-Guzik}},\ and\ \bibinfo {author}
  {\bibfnamefont {J.~L.}\ \bibnamefont {O'Brien}},\ }\bibfield  {title}
  {\bibinfo {title} {{A variational eigenvalue solver on a photonic quantum
  processor}},\ }\href {https://doi.org/10.1038/ncomms5213} {\bibfield
  {journal} {\bibinfo  {journal} {Nature Communications}\ }\textbf {\bibinfo
  {volume} {5}},\ \bibinfo {pages} {4213} (\bibinfo {year} {2014})},\ \Eprint
  {https://arxiv.org/abs/1304.3061} {arXiv:1304.3061} \BibitemShut {NoStop}%
\bibitem [{\citenamefont {McClean}\ \emph {et~al.}(2016)\citenamefont
  {McClean}, \citenamefont {Romero}, \citenamefont {Babbush},\ and\
  \citenamefont {Aspuru-Guzik}}]{McClean2016}%
  \BibitemOpen
  \bibfield  {author} {\bibinfo {author} {\bibfnamefont {J.~R.}\ \bibnamefont
  {McClean}}, \bibinfo {author} {\bibfnamefont {J.}~\bibnamefont {Romero}},
  \bibinfo {author} {\bibfnamefont {R.}~\bibnamefont {Babbush}},\ and\ \bibinfo
  {author} {\bibfnamefont {A.}~\bibnamefont {Aspuru-Guzik}},\ }\bibfield
  {title} {\bibinfo {title} {{The theory of variational hybrid
  quantum-classical algorithms}},\ }\href
  {https://doi.org/10.1088/1367-2630/18/2/023023} {\bibfield  {journal}
  {\bibinfo  {journal} {New Journal of Physics}\ }\textbf {\bibinfo {volume}
  {18}},\ \bibinfo {pages} {023023} (\bibinfo {year} {2016})},\ \Eprint
  {https://arxiv.org/abs/1509.04279} {arXiv:1509.04279} \BibitemShut {NoStop}%
\bibitem [{\citenamefont {Romero}\ \emph {et~al.}(2019)\citenamefont {Romero},
  \citenamefont {Babbush}, \citenamefont {McClean}, \citenamefont {Hempel},
  \citenamefont {Love},\ and\ \citenamefont {Aspuru-Guzik}}]{Romero2017}%
  \BibitemOpen
  \bibfield  {author} {\bibinfo {author} {\bibfnamefont {J.}~\bibnamefont
  {Romero}}, \bibinfo {author} {\bibfnamefont {R.}~\bibnamefont {Babbush}},
  \bibinfo {author} {\bibfnamefont {J.~R.}\ \bibnamefont {McClean}}, \bibinfo
  {author} {\bibfnamefont {C.}~\bibnamefont {Hempel}}, \bibinfo {author}
  {\bibfnamefont {P.~J.}\ \bibnamefont {Love}},\ and\ \bibinfo {author}
  {\bibfnamefont {A.}~\bibnamefont {Aspuru-Guzik}},\ }\bibfield  {title}
  {\bibinfo {title} {{Strategies for quantum computing molecular energies using
  the unitary coupled cluster ansatz}},\ }\href
  {https://doi.org/10.1088/2058-9565/aad3e4} {\bibfield  {journal} {\bibinfo
  {journal} {Quantum Science and Technology}\ }\textbf {\bibinfo {volume}
  {4}},\ \bibinfo {pages} {1168} (\bibinfo {year} {2019})},\ \Eprint
  {https://arxiv.org/abs/1701.02691} {arXiv:1701.02691} \BibitemShut {NoStop}%
\bibitem [{\citenamefont {Mazziotti}(1998{\natexlab{a}})}]{Mazziotti1998}%
  \BibitemOpen
  \bibfield  {author} {\bibinfo {author} {\bibfnamefont {D.~A.}\ \bibnamefont
  {Mazziotti}},\ }\bibfield  {title} {\bibinfo {title} {{Contracted
  Schr{\"{o}}dinger equation: Determining quantum energies and two-particle
  density matrices without wave functions}},\ }\href
  {https://doi.org/10.1103/PhysRevA.57.4219} {\bibfield  {journal} {\bibinfo
  {journal} {Physical Review A}\ }\textbf {\bibinfo {volume} {57}},\ \bibinfo
  {pages} {4219} (\bibinfo {year} {1998}{\natexlab{a}})}\BibitemShut {NoStop}%
\bibitem [{\citenamefont
  {Mazziotti}(1998{\natexlab{b}})}]{Mazziotti1998_schwinger}%
  \BibitemOpen
  \bibfield  {author} {\bibinfo {author} {\bibfnamefont {D.~a.}\ \bibnamefont
  {Mazziotti}},\ }\bibfield  {title} {\bibinfo {title} {{Approximate solution
  for electron correlation through the use of Schwinger probes}},\ }\href
  {https://doi.org/10.1016/S0009-2614(98)00470-9} {\bibfield  {journal}
  {\bibinfo  {journal} {Chemical Physics Letters}\ }\textbf {\bibinfo {volume}
  {289}},\ \bibinfo {pages} {419} (\bibinfo {year}
  {1998}{\natexlab{b}})}\BibitemShut {NoStop}%
\bibitem [{\citenamefont {Schlimgen}\ \emph {et~al.}(2021)\citenamefont
  {Schlimgen}, \citenamefont {Head-Marsden}, \citenamefont {Sager},
  \citenamefont {Narang},\ and\ \citenamefont {Mazziotti}}]{Schlimgen2021}%
  \BibitemOpen
  \bibfield  {author} {\bibinfo {author} {\bibfnamefont {A.~W.}\ \bibnamefont
  {Schlimgen}}, \bibinfo {author} {\bibfnamefont {K.}~\bibnamefont
  {Head-Marsden}}, \bibinfo {author} {\bibfnamefont {L.~M.}\ \bibnamefont
  {Sager}}, \bibinfo {author} {\bibfnamefont {P.}~\bibnamefont {Narang}},\ and\
  \bibinfo {author} {\bibfnamefont {D.~A.}\ \bibnamefont {Mazziotti}},\
  }\bibfield  {title} {\bibinfo {title} {{Quantum Simulation of Open Quantum
  Systems Using a Unitary Decomposition of Operators}},\ }\href
  {https://doi.org/10.1103/PhysRevLett.127.270503} {\bibfield  {journal}
  {\bibinfo  {journal} {Physical Review Letters}\ }\textbf {\bibinfo {volume}
  {127}},\ \bibinfo {pages} {270503} (\bibinfo {year} {2021})},\ \Eprint
  {https://arxiv.org/abs/2106.12588} {arXiv:2106.12588} \BibitemShut {NoStop}%
\bibitem [{\citenamefont {Nocedal}\ and\ \citenamefont
  {Wright}(2006)}]{Robinson2006}%
  \BibitemOpen
  \bibfield  {author} {\bibinfo {author} {\bibfnamefont {J.}~\bibnamefont
  {Nocedal}}\ and\ \bibinfo {author} {\bibfnamefont {S.~J.}\ \bibnamefont
  {Wright}},\ }\href {https://doi.org/10.1007/978-0-387-40065-5} {\emph
  {\bibinfo {title} {Numerical Optimization}}},\ Springer Series in Operations
  Research and Financial Engineering\ (\bibinfo  {publisher} {Springer New
  York},\ \bibinfo {year} {2006})\BibitemShut {NoStop}%
\bibitem [{\citenamefont {Smart}\ and\ \citenamefont
  {Mazziotti}(2022{\natexlab{a}})}]{Smart2022_opt}%
  \BibitemOpen
  \bibfield  {author} {\bibinfo {author} {\bibfnamefont {S.~E.}\ \bibnamefont
  {Smart}}\ and\ \bibinfo {author} {\bibfnamefont {D.~A.}\ \bibnamefont
  {Mazziotti}},\ }\bibfield  {title} {\bibinfo {title} {{Exploring Convergence
  of Quantum Contracted Eigensolvers through a Locally Parameterized
  Optimization}},\ }\href@noop {} {\bibfield  {journal} {\bibinfo  {journal}
  {Unpublished work}\ } (\bibinfo {year} {2022}{\natexlab{a}})}\BibitemShut
  {NoStop}%
\bibitem [{\citenamefont {Mazziotti}(2007{\natexlab{c}})}]{Mazziotti2007_book}%
  \BibitemOpen
  \bibinfo {editor} {\bibfnamefont {D.~A.}\ \bibnamefont {Mazziotti}},\ ed.,\
  \href {https://doi.org/10.1002/0470106603} {\emph {\bibinfo {title} {Advances
  in Chemical Physics}}},\ \bibinfo {series} {Advances in Chemical Physics},
  Vol.\ \bibinfo {volume} {134}\ (\bibinfo  {publisher} {John Wiley \& Sons,
  Inc.},\ \bibinfo {address} {Hoboken, NJ, USA},\ \bibinfo {year} {2007})\ p.\
  \bibinfo {pages} {574}\BibitemShut {NoStop}%
\bibitem [{\citenamefont {Nakatsuji}(1976)}]{Nakatsuji1976}%
  \BibitemOpen
  \bibfield  {author} {\bibinfo {author} {\bibfnamefont {H.}~\bibnamefont
  {Nakatsuji}},\ }\bibfield  {title} {\bibinfo {title} {{Equation for the
  direct determination of the density matrix}},\ }\href
  {https://doi.org/10.1103/PhysRevA.14.41} {\bibfield  {journal} {\bibinfo
  {journal} {Physical Review A}\ }\textbf {\bibinfo {volume} {14}},\ \bibinfo
  {pages} {41} (\bibinfo {year} {1976})}\BibitemShut {NoStop}%
\bibitem [{\citenamefont {{Hehre, W. J.; Ditchfield, R.;
  Pople}}(1972)}]{HehreW.J.;DitchfieldR.;Pople1972}%
  \BibitemOpen
  \bibfield  {author} {\bibinfo {author} {\bibfnamefont {J.~A.}\ \bibnamefont
  {{Hehre, W. J.; Ditchfield, R.; Pople}}},\ }\bibfield  {title} {\bibinfo
  {title} {{Self Consistent Molecular Orbital Methods. XII. Further Extensions
  of Gaussian Type Basis Sets for Use in Molecular Orbital Studies of Organic
  Molecules}},\ }\href {https://doi.org/10.1063/1.1677527} {\bibfield
  {journal} {\bibinfo  {journal} {J. Chem. Phys.}\ }\textbf {\bibinfo {volume}
  {56}},\ \bibinfo {pages} {2257} (\bibinfo {year} {1972})}\BibitemShut
  {NoStop}%
\bibitem [{\citenamefont {Coleman}(1997)}]{Coleman1997}%
  \BibitemOpen
  \bibfield  {author} {\bibinfo {author} {\bibfnamefont {A.~J.}\ \bibnamefont
  {Coleman}},\ }\bibfield  {title} {\bibinfo {title} {{The AGP model for
  fermion systems}},\ }\href
  {https://doi.org/10.1002/(SICI)1097-461X(1997)63:1<23::AID-QUA5>3.0.CO;2-4}
  {\bibfield  {journal} {\bibinfo  {journal} {International Journal of Quantum
  Chemistry}\ }\textbf {\bibinfo {volume} {63}},\ \bibinfo {pages} {23}
  (\bibinfo {year} {1997})}\BibitemShut {NoStop}%
\bibitem [{\citenamefont {Coleman}\ and\ \citenamefont
  {Yukalov}(2000)}]{Coleman2000}%
  \BibitemOpen
  \bibfield  {author} {\bibinfo {author} {\bibfnamefont {A.}~\bibnamefont
  {Coleman}}\ and\ \bibinfo {author} {\bibfnamefont {V.}~\bibnamefont
  {Yukalov}},\ }\href@noop {} {\emph {\bibinfo {title} {{Reduced Density
  Matrices: Coulson's Challenge}}}}\ (\bibinfo  {publisher} {Springer},\
  \bibinfo {address} {Berlin Heidelberg New York},\ \bibinfo {year}
  {2000})\BibitemShut {NoStop}%
\bibitem [{\citenamefont {Johnson}\ \emph {et~al.}(2013)\citenamefont
  {Johnson}, \citenamefont {Ayers}, \citenamefont {Limacher}, \citenamefont
  {Baerdemacker}, \citenamefont {Neck},\ and\ \citenamefont
  {Bultinck}}]{Johnson2013}%
  \BibitemOpen
  \bibfield  {author} {\bibinfo {author} {\bibfnamefont {P.~A.}\ \bibnamefont
  {Johnson}}, \bibinfo {author} {\bibfnamefont {P.~W.}\ \bibnamefont {Ayers}},
  \bibinfo {author} {\bibfnamefont {P.~A.}\ \bibnamefont {Limacher}}, \bibinfo
  {author} {\bibfnamefont {S.~D.}\ \bibnamefont {Baerdemacker}}, \bibinfo
  {author} {\bibfnamefont {D.~V.}\ \bibnamefont {Neck}},\ and\ \bibinfo
  {author} {\bibfnamefont {P.}~\bibnamefont {Bultinck}},\ }\bibfield  {title}
  {\bibinfo {title} {{A size-consistent approach to strongly correlated systems
  using a generalized antisymmetrized product of nonorthogonal geminals}},\
  }\href {https://doi.org/10.1016/j.comptc.2012.09.030} {\bibfield  {journal}
  {\bibinfo  {journal} {Computational and Theoretical Chemistry}\ }\textbf
  {\bibinfo {volume} {1003}},\ \bibinfo {pages} {101} (\bibinfo {year}
  {2013})}\BibitemShut {NoStop}%
\bibitem [{\citenamefont {Stein}\ \emph {et~al.}(2014)\citenamefont {Stein},
  \citenamefont {Henderson},\ and\ \citenamefont {Scuseria}}]{Stein2014}%
  \BibitemOpen
  \bibfield  {author} {\bibinfo {author} {\bibfnamefont {T.}~\bibnamefont
  {Stein}}, \bibinfo {author} {\bibfnamefont {T.~M.}\ \bibnamefont
  {Henderson}},\ and\ \bibinfo {author} {\bibfnamefont {G.~E.}\ \bibnamefont
  {Scuseria}},\ }\bibfield  {title} {\bibinfo {title} {{Seniority zero pair
  coupled cluster doubles theory}},\ }\bibfield  {journal} {\bibinfo  {journal}
  {Journal of Chemical Physics}\ }\textbf {\bibinfo {volume} {140}},\ \href
  {https://doi.org/10.1063/1.4880819} {10.1063/1.4880819} (\bibinfo {year}
  {2014})\BibitemShut {NoStop}%
\bibitem [{\citenamefont {Sager}\ and\ \citenamefont
  {Mazziotti}(2022)}]{Sager2022}%
  \BibitemOpen
  \bibfield  {author} {\bibinfo {author} {\bibfnamefont {L.~A.~M.}\
  \bibnamefont {Sager}}\ and\ \bibinfo {author} {\bibfnamefont {D.~A.}\
  \bibnamefont {Mazziotti}},\ }\bibfield  {title} {\bibinfo {title}
  {{Cooper-pair condensates with nonclassical long-range order on quantum
  devices}},\ }\href {https://doi.org/10.1103/PhysRevResearch.4.013003}
  {\bibfield  {journal} {\bibinfo  {journal} {Physical Review Research}\
  }\textbf {\bibinfo {volume} {4}},\ \bibinfo {pages} {1} (\bibinfo {year}
  {2022})}\BibitemShut {NoStop}%
\bibitem [{\citenamefont {Bonet-Monroig}\ \emph {et~al.}(2020)\citenamefont
  {Bonet-Monroig}, \citenamefont {Babbush},\ and\ \citenamefont
  {O'Brien}}]{Bonet-Monroig2019}%
  \BibitemOpen
  \bibfield  {author} {\bibinfo {author} {\bibfnamefont {X.}~\bibnamefont
  {Bonet-Monroig}}, \bibinfo {author} {\bibfnamefont {R.}~\bibnamefont
  {Babbush}},\ and\ \bibinfo {author} {\bibfnamefont {T.~E.}\ \bibnamefont
  {O'Brien}},\ }\bibfield  {title} {\bibinfo {title} {{Nearly Optimal
  Measurement Scheduling for Partial Tomography of Quantum States}},\ }\href
  {https://doi.org/10.1103/PhysRevX.10.031064} {\bibfield  {journal} {\bibinfo
  {journal} {Physical Review X}\ }\textbf {\bibinfo {volume} {10}},\ \bibinfo
  {pages} {031064} (\bibinfo {year} {2020})},\ \Eprint
  {https://arxiv.org/abs/1908.05628} {arXiv:1908.05628} \BibitemShut {NoStop}%
\bibitem [{\citenamefont {Gokhale}\ and\ \citenamefont
  {Chong}(2019)}]{Gokhale2019}%
  \BibitemOpen
  \bibfield  {author} {\bibinfo {author} {\bibfnamefont {P.}~\bibnamefont
  {Gokhale}}\ and\ \bibinfo {author} {\bibfnamefont {F.~T.}\ \bibnamefont
  {Chong}},\ }\bibfield  {title} {\bibinfo {title} {{$O(N^3)$ Measurement Cost
  for Variational Quantum Eigensolver on Molecular Hamiltonians}},\ }\href
  {http://arxiv.org/abs/1908.11857} {\bibfield  {journal} {\bibinfo  {journal}
  {arXiv}\ } (\bibinfo {year} {2019})},\ \Eprint
  {https://arxiv.org/abs/1908.11857} {arXiv:1908.11857} \BibitemShut {NoStop}%
\bibitem [{\citenamefont {Zhao}\ \emph {et~al.}(2020)\citenamefont {Zhao},
  \citenamefont {Rubin},\ and\ \citenamefont {Miyake}}]{Zhao2020}%
  \BibitemOpen
  \bibfield  {author} {\bibinfo {author} {\bibfnamefont {A.}~\bibnamefont
  {Zhao}}, \bibinfo {author} {\bibfnamefont {N.~C.}\ \bibnamefont {Rubin}},\
  and\ \bibinfo {author} {\bibfnamefont {A.}~\bibnamefont {Miyake}},\
  }\bibfield  {title} {\bibinfo {title} {{Fermionic partial tomography via
  classical shadows}},\ }\href {https://doi.org/10.1103/PhysRevLett.127.110504}
  {\bibfield  {journal} {\bibinfo  {journal} {Physical Review Letters}\
  }\textbf {\bibinfo {volume} {127}},\ \bibinfo {pages} {110504} (\bibinfo
  {year} {2020})},\ \Eprint {https://arxiv.org/abs/2010.16094}
  {arXiv:2010.16094} \BibitemShut {NoStop}%
\bibitem [{\citenamefont {Smart}\ and\ \citenamefont
  {Mazziotti}(2021{\natexlab{b}})}]{Smart2021_tomo}%
  \BibitemOpen
  \bibfield  {author} {\bibinfo {author} {\bibfnamefont {S.~E.}\ \bibnamefont
  {Smart}}\ and\ \bibinfo {author} {\bibfnamefont {D.~A.}\ \bibnamefont
  {Mazziotti}},\ }\bibfield  {title} {\bibinfo {title} {{Lowering tomography
  costs in quantum simulation with a symmetry projected operator basis}},\
  }\href {https://doi.org/10.1103/PhysRevA.103.012420} {\bibfield  {journal}
  {\bibinfo  {journal} {Physical Review A}\ }\textbf {\bibinfo {volume}
  {103}},\ \bibinfo {pages} {012420} (\bibinfo {year} {2021}{\natexlab{b}})},\
  \Eprint {https://arxiv.org/abs/2008.06027} {arXiv:2008.06027} \BibitemShut
  {NoStop}%
\bibitem [{\citenamefont {McClean}\ \emph {et~al.}(2018)\citenamefont
  {McClean}, \citenamefont {Boixo}, \citenamefont {Smelyanskiy}, \citenamefont
  {Babbush},\ and\ \citenamefont {Neven}}]{McClean2018}%
  \BibitemOpen
  \bibfield  {author} {\bibinfo {author} {\bibfnamefont {J.~R.}\ \bibnamefont
  {McClean}}, \bibinfo {author} {\bibfnamefont {S.}~\bibnamefont {Boixo}},
  \bibinfo {author} {\bibfnamefont {V.~N.}\ \bibnamefont {Smelyanskiy}},
  \bibinfo {author} {\bibfnamefont {R.}~\bibnamefont {Babbush}},\ and\ \bibinfo
  {author} {\bibfnamefont {H.}~\bibnamefont {Neven}},\ }\bibfield  {title}
  {\bibinfo {title} {{Barren plateaus in quantum neural network training
  landscapes}},\ }\href {https://doi.org/10.1038/s41467-018-07090-4} {\bibfield
   {journal} {\bibinfo  {journal} {Nature Communications}\ }\textbf {\bibinfo
  {volume} {9}},\ \bibinfo {pages} {1} (\bibinfo {year} {2018})},\ \Eprint
  {https://arxiv.org/abs/1803.11173} {arXiv:1803.11173} \BibitemShut {NoStop}%
\bibitem [{\citenamefont {Yordanov}\ \emph {et~al.}(2021)\citenamefont
  {Yordanov}, \citenamefont {Armaos}, \citenamefont {Barnes},\ and\
  \citenamefont {Arvidsson-Shukur}}]{Yordanov2021}%
  \BibitemOpen
  \bibfield  {author} {\bibinfo {author} {\bibfnamefont {Y.~S.}\ \bibnamefont
  {Yordanov}}, \bibinfo {author} {\bibfnamefont {V.}~\bibnamefont {Armaos}},
  \bibinfo {author} {\bibfnamefont {C.~H.}\ \bibnamefont {Barnes}},\ and\
  \bibinfo {author} {\bibfnamefont {D.~R.}\ \bibnamefont {Arvidsson-Shukur}},\
  }\bibfield  {title} {\bibinfo {title} {{Qubit-excitation-based adaptive
  variational quantum eigensolver}},\ }\href
  {https://doi.org/10.1038/s42005-021-00730-0} {\bibfield  {journal} {\bibinfo
  {journal} {Communications Physics}\ }\textbf {\bibinfo {volume} {4}},\
  \bibinfo {pages} {1} (\bibinfo {year} {2021})},\ \Eprint
  {https://arxiv.org/abs/2011.10540} {arXiv:2011.10540} \BibitemShut {NoStop}%
\bibitem [{\citenamefont {Berry}\ \emph {et~al.}(2015)\citenamefont {Berry},
  \citenamefont {Childs},\ and\ \citenamefont {Kothari}}]{Berry2015}%
  \BibitemOpen
  \bibfield  {author} {\bibinfo {author} {\bibfnamefont {D.~W.}\ \bibnamefont
  {Berry}}, \bibinfo {author} {\bibfnamefont {A.~M.}\ \bibnamefont {Childs}},\
  and\ \bibinfo {author} {\bibfnamefont {R.}~\bibnamefont {Kothari}},\
  }\bibfield  {title} {\bibinfo {title} {{Hamiltonian Simulation with Nearly
  Optimal Dependence on all Parameters}},\ }\href
  {https://doi.org/10.1109/FOCS.2015.54} {\bibfield  {journal} {\bibinfo
  {journal} {Proceedings - Annual IEEE Symposium on Foundations of Computer
  Science, FOCS}\ }\textbf {\bibinfo {volume} {2015-Decem}},\ \bibinfo {pages}
  {792} (\bibinfo {year} {2015})},\ \Eprint {https://arxiv.org/abs/1501.01715}
  {arXiv:1501.01715} \BibitemShut {NoStop}%
\bibitem [{\citenamefont {Low}\ and\ \citenamefont {Chuang}(2019)}]{Low2019}%
  \BibitemOpen
  \bibfield  {author} {\bibinfo {author} {\bibfnamefont {G.~H.}\ \bibnamefont
  {Low}}\ and\ \bibinfo {author} {\bibfnamefont {I.~L.}\ \bibnamefont
  {Chuang}},\ }\bibfield  {title} {\bibinfo {title} {{Hamiltonian Simulation by
  Qubitization}},\ }\href {https://doi.org/10.22331/q-2019-07-12-163}
  {\bibfield  {journal} {\bibinfo  {journal} {Quantum}\ }\textbf {\bibinfo
  {volume} {3}},\ \bibinfo {pages} {163} (\bibinfo {year} {2019})},\ \Eprint
  {https://arxiv.org/abs/1610.06546} {arXiv:1610.06546} \BibitemShut {NoStop}%
\bibitem [{\citenamefont {Lemieux}\ \emph {et~al.}(2020)\citenamefont
  {Lemieux}, \citenamefont {Heim}, \citenamefont {Poulin}, \citenamefont
  {Svore},\ and\ \citenamefont {Troyer}}]{Lemieux2020}%
  \BibitemOpen
  \bibfield  {author} {\bibinfo {author} {\bibfnamefont {J.}~\bibnamefont
  {Lemieux}}, \bibinfo {author} {\bibfnamefont {B.}~\bibnamefont {Heim}},
  \bibinfo {author} {\bibfnamefont {D.}~\bibnamefont {Poulin}}, \bibinfo
  {author} {\bibfnamefont {K.}~\bibnamefont {Svore}},\ and\ \bibinfo {author}
  {\bibfnamefont {M.}~\bibnamefont {Troyer}},\ }\bibfield  {title} {\bibinfo
  {title} {{Efficient quantum walk circuits for metropolis-hastings
  algorithm}},\ }\bibfield  {journal} {\bibinfo  {journal} {Quantum}\ }\textbf
  {\bibinfo {volume} {4}},\ \href {https://doi.org/10.22331/Q-2020-06-29-287}
  {10.22331/Q-2020-06-29-287} (\bibinfo {year} {2020}),\ \Eprint
  {https://arxiv.org/abs/1910.01659} {arXiv:1910.01659} \BibitemShut {NoStop}%
\bibitem [{\citenamefont {Beebe}\ and\ \citenamefont
  {Linderberg}(1977)}]{Beebe1977}%
  \BibitemOpen
  \bibfield  {author} {\bibinfo {author} {\bibfnamefont {N.~H.~F.}\
  \bibnamefont {Beebe}}\ and\ \bibinfo {author} {\bibfnamefont
  {J.}~\bibnamefont {Linderberg}},\ }\bibfield  {title} {\bibinfo {title}
  {{Simplifications in the Two-Electron Integral Array in Molecular
  Calculations}},\ }\href {https://doi.org/10.1002/qua.560120408} {\bibfield
  {journal} {\bibinfo  {journal} {Int. J. Quant. Chem.}\ }\textbf {\bibinfo
  {volume} {12}},\ \bibinfo {pages} {683} (\bibinfo {year} {1977})}\BibitemShut
  {NoStop}%
\bibitem [{\citenamefont {Motta}\ \emph {et~al.}(2021)\citenamefont {Motta},
  \citenamefont {Ye}, \citenamefont {McClean}, \citenamefont {Li},
  \citenamefont {Minnich}, \citenamefont {Babbush},\ and\ \citenamefont
  {Chan}}]{Motta2018}%
  \BibitemOpen
  \bibfield  {author} {\bibinfo {author} {\bibfnamefont {M.}~\bibnamefont
  {Motta}}, \bibinfo {author} {\bibfnamefont {E.}~\bibnamefont {Ye}}, \bibinfo
  {author} {\bibfnamefont {J.~R.}\ \bibnamefont {McClean}}, \bibinfo {author}
  {\bibfnamefont {Z.}~\bibnamefont {Li}}, \bibinfo {author} {\bibfnamefont
  {A.~J.}\ \bibnamefont {Minnich}}, \bibinfo {author} {\bibfnamefont
  {R.}~\bibnamefont {Babbush}},\ and\ \bibinfo {author} {\bibfnamefont
  {G.~K.-L.}\ \bibnamefont {Chan}},\ }\bibfield  {title} {\bibinfo {title}
  {{Low rank representations for quantum simulation of electronic structure}},\
  }\href {https://doi.org/10.1038/s41534-021-00416-z} {\bibfield  {journal}
  {\bibinfo  {journal} {npj Quantum Information}\ }\textbf {\bibinfo {volume}
  {7}},\ \bibinfo {pages} {83} (\bibinfo {year} {2021})},\ \Eprint
  {https://arxiv.org/abs/1808.02625} {arXiv:1808.02625} \BibitemShut {NoStop}%
\bibitem [{\citenamefont {Hohenstein}\ and\ \citenamefont
  {Sherrill}(2010)}]{Hohenstein2010}%
  \BibitemOpen
  \bibfield  {author} {\bibinfo {author} {\bibfnamefont {E.~G.}\ \bibnamefont
  {Hohenstein}}\ and\ \bibinfo {author} {\bibfnamefont {C.~D.}\ \bibnamefont
  {Sherrill}},\ }\bibfield  {title} {\bibinfo {title} {{Density fitting and
  Cholesky decomposition approximations in symmetry-adapted perturbation
  theory: Implementation and application to probe the nature of $\pi$ -
  $\pi$}},\ }\bibfield  {journal} {\bibinfo  {journal} {Journal of Chemical
  Physics}\ }\textbf {\bibinfo {volume} {132}},\ \href
  {https://doi.org/10.1063/1.3426316} {10.1063/1.3426316} (\bibinfo {year}
  {2010})\BibitemShut {NoStop}%
\bibitem [{\citenamefont {Kivlichan}\ \emph {et~al.}(2018)\citenamefont
  {Kivlichan}, \citenamefont {McClean}, \citenamefont {Wiebe}, \citenamefont
  {Gidney}, \citenamefont {Aspuru-Guzik}, \citenamefont {Chan},\ and\
  \citenamefont {Babbush}}]{Kivlichan2018}%
  \BibitemOpen
  \bibfield  {author} {\bibinfo {author} {\bibfnamefont {I.~D.}\ \bibnamefont
  {Kivlichan}}, \bibinfo {author} {\bibfnamefont {J.}~\bibnamefont {McClean}},
  \bibinfo {author} {\bibfnamefont {N.}~\bibnamefont {Wiebe}}, \bibinfo
  {author} {\bibfnamefont {C.}~\bibnamefont {Gidney}}, \bibinfo {author}
  {\bibfnamefont {A.}~\bibnamefont {Aspuru-Guzik}}, \bibinfo {author}
  {\bibfnamefont {G.~K.~L.}\ \bibnamefont {Chan}},\ and\ \bibinfo {author}
  {\bibfnamefont {R.}~\bibnamefont {Babbush}},\ }\bibfield  {title} {\bibinfo
  {title} {{Quantum Simulation of Electronic Structure with Linear Depth and
  Connectivity}},\ }\href {https://doi.org/10.1103/PhysRevLett.120.110501}
  {\bibfield  {journal} {\bibinfo  {journal} {Physical Review Letters}\
  }\textbf {\bibinfo {volume} {120}},\ \bibinfo {pages} {110501} (\bibinfo
  {year} {2018})},\ \Eprint {https://arxiv.org/abs/1711.04789}
  {arXiv:1711.04789} \BibitemShut {NoStop}%
\bibitem [{\citenamefont {Harbrecht}\ \emph {et~al.}(2012)\citenamefont
  {Harbrecht}, \citenamefont {Peters},\ and\ \citenamefont
  {Schneider}}]{HARBRECHT2012428}%
  \BibitemOpen
  \bibfield  {author} {\bibinfo {author} {\bibfnamefont {H.}~\bibnamefont
  {Harbrecht}}, \bibinfo {author} {\bibfnamefont {M.}~\bibnamefont {Peters}},\
  and\ \bibinfo {author} {\bibfnamefont {R.}~\bibnamefont {Schneider}},\
  }\bibfield  {title} {\bibinfo {title} {On the low-rank approximation by the
  pivoted cholesky decomposition},\ }\href
  {https://doi.org/https://doi.org/10.1016/j.apnum.2011.10.001} {\bibfield
  {journal} {\bibinfo  {journal} {Applied Numerical Mathematics}\ }\textbf
  {\bibinfo {volume} {62}},\ \bibinfo {pages} {428} (\bibinfo {year} {2012})},\
  \bibinfo {note} {third Chilean Workshop on Numerical Analysis of Partial
  Differential Equations (WONAPDE 2010)}\BibitemShut {NoStop}%
\bibitem [{\citenamefont {Smart}\ and\ \citenamefont
  {Mazziotti}(2022{\natexlab{b}})}]{Smart_hqca__hybrid}%
  \BibitemOpen
  \bibfield  {author} {\bibinfo {author} {\bibfnamefont {S.~E.}\ \bibnamefont
  {Smart}}\ and\ \bibinfo {author} {\bibfnamefont {D.~A.}\ \bibnamefont
  {Mazziotti}},\ }\href {https://github.com/damazz/HQCA} {\bibinfo {title}
  {{hqca - hybrid quantum computing algorithms for quantum chemistry}}}
  (\bibinfo {year} {2022}{\natexlab{b}})\BibitemShut {NoStop}%
\bibitem [{\citenamefont {Abraham}\ \emph {et~al.}(2019)\citenamefont
  {Abraham}, \citenamefont {AduOffei}, \citenamefont {Agarwal}, \citenamefont
  {Akhalwaya}, \citenamefont {Aleksandrowicz}, \citenamefont {Alexander},
  \citenamefont {Amy}, \citenamefont {Arbel}, \citenamefont {Arijit02},
  \citenamefont {Asfaw}, \citenamefont {Avkhadiev}, \citenamefont {Azaustre},
  \citenamefont {AzizNgoueya}, \citenamefont {Banerjee}, \citenamefont
  {Bansal}, \citenamefont {Barkoutsos}, \citenamefont {Barron}, \citenamefont
  {Barron}, \citenamefont {Bello}, \citenamefont {Ben-Haim}, \citenamefont
  {Bevenius}, \citenamefont {Bhobe}, \citenamefont {Bishop} \emph
  {et~al.}}]{Qiskit}%
  \BibitemOpen
  \bibfield  {author} {\bibinfo {author} {\bibfnamefont {H.}~\bibnamefont
  {Abraham}}, \bibinfo {author} {\bibnamefont {AduOffei}}, \bibinfo {author}
  {\bibfnamefont {R.}~\bibnamefont {Agarwal}}, \bibinfo {author} {\bibfnamefont
  {I.~Y.}\ \bibnamefont {Akhalwaya}}, \bibinfo {author} {\bibfnamefont
  {G.}~\bibnamefont {Aleksandrowicz}}, \bibinfo {author} {\bibfnamefont
  {T.}~\bibnamefont {Alexander}}, \bibinfo {author} {\bibfnamefont
  {M.}~\bibnamefont {Amy}}, \bibinfo {author} {\bibfnamefont {E.}~\bibnamefont
  {Arbel}}, \bibinfo {author} {\bibnamefont {Arijit02}}, \bibinfo {author}
  {\bibfnamefont {A.}~\bibnamefont {Asfaw}}, \bibinfo {author} {\bibfnamefont
  {A.}~\bibnamefont {Avkhadiev}}, \bibinfo {author} {\bibfnamefont
  {C.}~\bibnamefont {Azaustre}}, \bibinfo {author} {\bibnamefont
  {AzizNgoueya}}, \bibinfo {author} {\bibfnamefont {A.}~\bibnamefont
  {Banerjee}}, \bibinfo {author} {\bibfnamefont {A.}~\bibnamefont {Bansal}},
  \bibinfo {author} {\bibfnamefont {P.}~\bibnamefont {Barkoutsos}}, \bibinfo
  {author} {\bibfnamefont {G.}~\bibnamefont {Barron}}, \bibinfo {author}
  {\bibfnamefont {G.~S.}\ \bibnamefont {Barron}}, \bibinfo {author}
  {\bibfnamefont {L.}~\bibnamefont {Bello}}, \bibinfo {author} {\bibfnamefont
  {Y.}~\bibnamefont {Ben-Haim}}, \bibinfo {author} {\bibfnamefont
  {D.}~\bibnamefont {Bevenius}}, \bibinfo {author} {\bibfnamefont
  {A.}~\bibnamefont {Bhobe}}, \bibinfo {author} {\bibfnamefont {L.~S.}\
  \bibnamefont {Bishop}}, \emph {et~al.},\ }\href
  {https://doi.org/10.5281/zenodo.2562110} {\bibinfo {title} {Qiskit: An
  open-source framework for quantum computing}} (\bibinfo {year}
  {2019})\BibitemShut {NoStop}%
\bibitem [{\citenamefont {Sun}\ \emph {et~al.}(2018)\citenamefont {Sun},
  \citenamefont {Berkelbach}, \citenamefont {Blunt}, \citenamefont {Booth},
  \citenamefont {Guo}, \citenamefont {Li}, \citenamefont {Liu}, \citenamefont
  {McClain}, \citenamefont {Sayfutyarova}, \citenamefont {Sharma},
  \citenamefont {Wouters},\ and\ \citenamefont {Chan}}]{PySCF}%
  \BibitemOpen
  \bibfield  {author} {\bibinfo {author} {\bibfnamefont {Q.}~\bibnamefont
  {Sun}}, \bibinfo {author} {\bibfnamefont {T.~C.}\ \bibnamefont {Berkelbach}},
  \bibinfo {author} {\bibfnamefont {N.~S.}\ \bibnamefont {Blunt}}, \bibinfo
  {author} {\bibfnamefont {G.~H.}\ \bibnamefont {Booth}}, \bibinfo {author}
  {\bibfnamefont {S.}~\bibnamefont {Guo}}, \bibinfo {author} {\bibfnamefont
  {Z.}~\bibnamefont {Li}}, \bibinfo {author} {\bibfnamefont {J.}~\bibnamefont
  {Liu}}, \bibinfo {author} {\bibfnamefont {J.~D.}\ \bibnamefont {McClain}},
  \bibinfo {author} {\bibfnamefont {E.~R.}\ \bibnamefont {Sayfutyarova}},
  \bibinfo {author} {\bibfnamefont {S.}~\bibnamefont {Sharma}}, \bibinfo
  {author} {\bibfnamefont {S.}~\bibnamefont {Wouters}},\ and\ \bibinfo {author}
  {\bibfnamefont {G.~K.-L.}\ \bibnamefont {Chan}},\ }\bibfield  {title}
  {\bibinfo {title} {Pyscf: the python-based simulations of chemistry
  framework},\ }\href {https://doi.org/https://doi.org/10.1002/wcms.1340}
  {\bibfield  {journal} {\bibinfo  {journal} {WIREs Computational Molecular
  Science}\ }\textbf {\bibinfo {volume} {8}},\ \bibinfo {pages} {e1340}
  (\bibinfo {year} {2018})},\ \Eprint
  {https://arxiv.org/abs/https://onlinelibrary.wiley.com/doi/pdf/10.1002/wcms.1340}
  {https://onlinelibrary.wiley.com/doi/pdf/10.1002/wcms.1340} \BibitemShut
  {NoStop}%
\end{thebibliography}%

\end{document}